\documentclass[pre, twocolumn, aps, longbibliography] {revtex4-2}

\usepackage{graphicx,epsfig}
\usepackage{amssymb, amsmath}
\usepackage{amsfonts}
\usepackage{color}
\usepackage{upgreek}
\usepackage[utf8]{inputenc}

\usepackage{algorithm}
\usepackage{algpseudocode}
\usepackage{array}

\usepackage{subcaption}
\usepackage[justification=raggedright,singlelinecheck=false]{caption}
\usepackage{footnote}
\usepackage{footmisc}

\newcommand \beq {\begin{equation}}
\newcommand \eeq {\end{equation}}
\newcommand \be {\begin{equation}}
\newcommand \ee {\end{equation}}

\newcommand \beqn {\begin{eqnarray}}
\newcommand \eeqn {\end{eqnarray}}

\begin{document}

\title{Asymmetric Opinion Formation of Emotional Eccitable Agents}


\author{Irene Ferri$^{1,2}$}
\email{irene.ferri@ub.edu}
\author{Emanuele Cozzo$^{1,2,3}$}
\author{Aleix Nicol\'as-Oliv\'e$^{1,2}$}
\author{Albert D\'iaz-Guilera$^{1,2}$}
\author{Luce Prignano$^{1,2}$}

\affiliation{$^1$Departament de F\'isica de la Mat\`eria Condensada, Universitat de Barcelona (UB), c. Mart\'i i Franqu\`es, 1, 08028 Barcelona, Spain}
\affiliation{$^2$Institut de Recerca en Sistemes Complexos (UBICS), Universitat de Barcelona (UB), Barcelona, Spain}
\affiliation{$^3$CNSc-IN3, Universitat Oberta de Catalunya, Barcelona, Spain}

\date{\today}

\begin{abstract}
{\bf Abstract:} 
The bounded confidence model represents a widely adopted framework for modeling opinion dynamics wherein actors have a continuous-valued opinion and interact and approach their positions in the opinion space only if their opinions are within a specified confidence threshold. Here, we propose a novel framework where the confidence bound is determined by a decreasing function of their emotional arousal, an additional independent variable distinct from the opinion value. Additionally, our framework accounts for agents' ability to broadcast messages, with interactions influencing the timing of each other's message emissions. Our findings underscore the significant role of synchronization in shaping consensus formation. Furthermore, we demonstrate that variable confidence intervals alter the impact of step length when navigating the opinion space, leading to deviations from observations in the traditional Deffuant model.

\end{abstract}

\maketitle

\section{Introduction} \label{Sec:Intro}

In opinion dynamics, formal and computational models describe the formation of opinion groups within human communities. A primary distinction in these models is between those that represent opinion as a discrete variable, such as the voter model \cite{liggett2013stochastic}, and those that represent it as a continuous variable. In the latter, opinions range continuously between two extreme values (usually $0$ and $1$), and agents adjust their opinions based on interactions with other agents \cite{lorenz2007continuous}. A common feature of continuous opinion dynamics models is the assumption of bounded confidence \cite{lorenz2007continuous}, which posits that agents only change their opinions when interacting with others whose opinions are not too different from their own, leading to a compromise.

Since the early 2000s, two models have garnered significant attention in the sociophysics community \cite{Castellano_2009}: the Krause and Hegselmann (KH) model \cite{Hegselmann_2002} and the Deffuant and Weisbuch (DW) model \cite{Deffuant_2000}. These models explore the link between micro-level processes of social influence and homophily and macro-level phenomena of consensus or fragmentation in human communities \cite{flache2017models}. Recently, there has been a consensus in the field that future research should pursue two main directions: developing theoretical frameworks that compare, relate, and integrate different models \cite{flache2017models} and grounding these models in empirical data \cite{flache2017models,carpentras2023we,sobkowicz2009modelling}. This work focuses on the first direction by establishing a general modeling framework for discrete-time continuous opinion models with bounded confidence while maintaining an empirical grounding. We integrate new micro-level processes supported by empirical evidence and propose a framework to connect our model to the empirical world.

In particular, we aim to connect to the field of digital democracy \cite{helbing2023democracy}. This new research field as emerged as a result of the increasing use of technological tools to implement participatory democratic processes. Research in opinion dynamics needs to develop formal and computational models general enough to understand these dynamics and assist in designing technopolitical networks \cite{barandiaran2024decidim} where people form opinions and make decisions. This work aims to contribute to that effort by presenting a technological tool that can serve as both an experimental device for calibrating the model and a technological aid for deliberative processes (see Appendix \ref{App: dialoguem}).

The main difference between the DW and HK models lies in the \textit{communication regimes} \cite{urbig2008opinion}, which determine the number of agents that interact and adjust opinions at each time step, essentially defining "who talks to whom." Different communication regimes model various communication situations. Both models are symmetric, meaning all interacting agents update their opinions. The impact of asymmetry between emitters and receivers on consensus formation is explored in \cite{veefkind2022consensus}. In the DW model, agents interact in randomly selected pairs, whereas, in the HK model, agents perceive all other agents. In \cite{urbig2008opinion} a generalization that interpolates between these extremes is proposed, showing that communication regimes strongly affect opinion dynamics. However, in all these formulations, communication regimes are external to the opinion dynamics and must be specified \textit{a priori}.

Our approach, instead, couples the evolution of communication regimes to the opinion dynamics. In studying information diffusion on social media, the emission of a message by an agent is often modeled as a threshold process \cite{piedrahita2013modeling}. Threshold models, common in social phenomena modeling, are well-suited for describing self-sustained activity and synchronization, which in communication manifests as information cascades. A step toward integrating this approach into opinion dynamics has been explored in \cite{Schweitzer_2020}, where agents are coupled to an information field and emit messages when a threshold is surpassed. Our work moves in the same direction by modeling individuals as agents with an internal propensity to emit their opinions, which can be influenced by interactions with other agents holding similar opinions, akin to modeling information cascades \cite{piedrahita2013modeling}, with a parameter controlling the transition between different communication regimes that can be inferred from experimental data and adapted to different communication situations.

Finally, in recent years, another trend in theoretical work on opinion dynamics is the explicit integration of emotional states in modeling micro-level processes affecting opinion formation \cite{Schweitzer_2020,Sobkowicz_2015,sobkowicz2012discrete}. Empirical observations often link the level of emotional arousal to an agent's confidence level. We integrate emotional arousal into our model as an additional dimension of the opinion space, allowing for the emergence of zealots through interactions \cite{colaiori2016consensus}.

Thus, from a physics perspective, our agents can be formally modeled as interacting integrate-and-fire oscillators moving in a two-dimensional plane with limited vision \cite{prignano2013tuning}. Agents accumulate influence from nearby opinions similarly to how oscillators integrate input signals until a threshold triggers a response, with the confidence bound being analogous to the concept of limited vision. Emergent collective behavior, such as consensus, polarization, or sustained opinion diversity, is driven by thresholds and limited interaction ranges in the same way as synchronization phenomena in physical systems.

The paper is organized as follows. In Section II, we introduce the model, detailing the interaction rules and dynamics governing the opinion and emotional arousal variables. Section III presents the results of our simulations, focusing on opinion group formation, synchronization effects, and the impact of varying control parameters. In Section IV, we discuss the implications of our findings in the context of consensus and polarization, particularly emphasizing the role of synchronization and emotional arousal in opinion fragmentation. Finally, Section V concludes the paper by summarizing the key insights and outlining potential future research directions in the field of opinion dynamics and technopolitical applications.
  
%
\section{The Model} \label{Sec:the_model}
\begin{figure}[h!]
\centering
\includegraphics[width=0.9\linewidth]{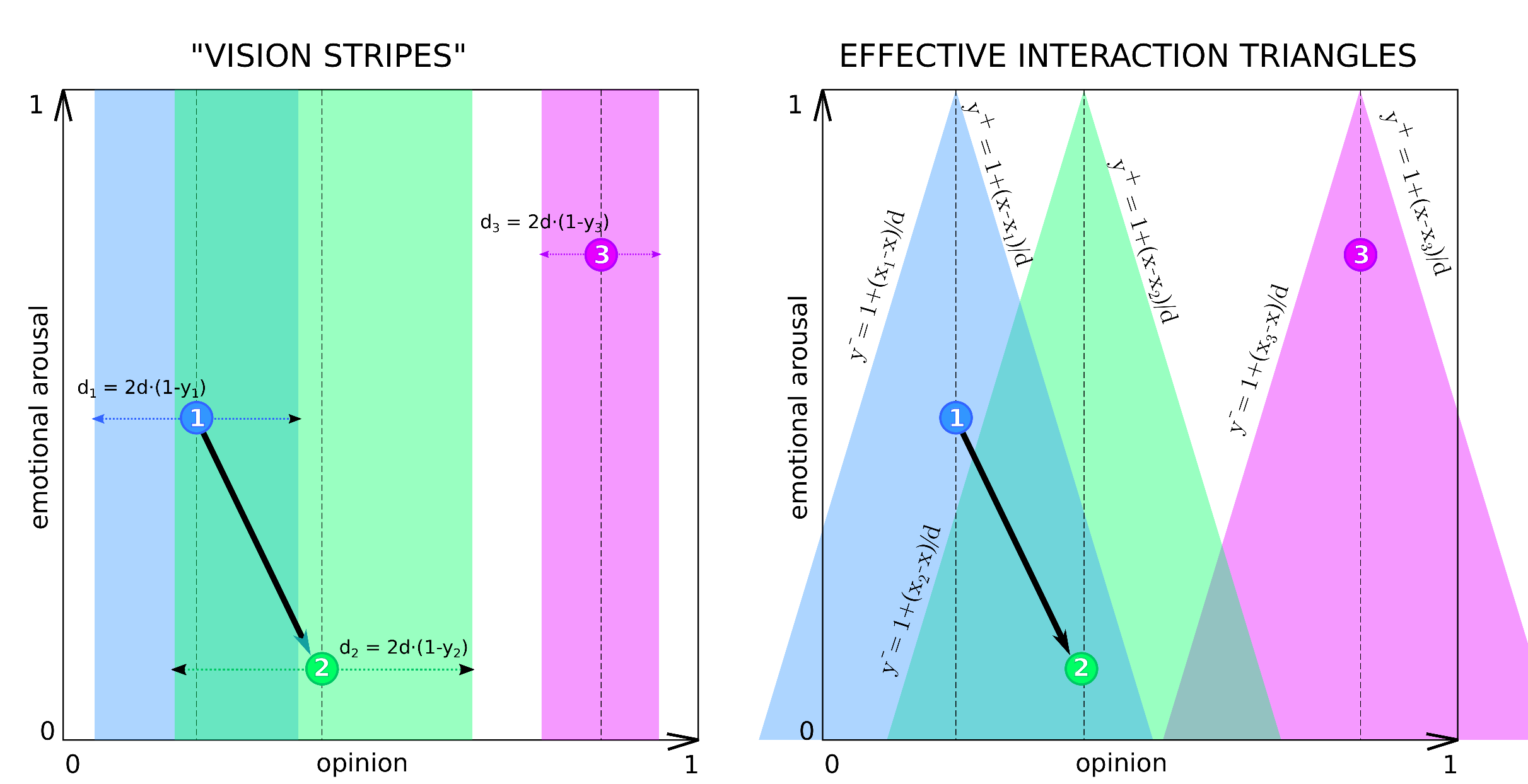}
\caption{(Left) Confidence area: the region within an emitter must be located for a receiver at a given opinion position $O_j$ to listen to their message. (Right) Affectation area: the region within which receivers will be affected by emitters at a given opinion position $O_i$.}
\label{Fig:bands}       
\end{figure}

Agents in our model are integrate-and-fire oscillators (IFOs). An IFO is a system that integrates a function until some threshold value is reached, at which point the system "fires" and the integrator is reset to zero\cite{bottani1996synchronization}, in this context the firing event represents a communicative action, i.e a message being send. We consider a population of $N$ agents, which are initially distributed uniformly at random on a plane, as represented in Fig. \ref{Fig:bands}. In line with the classical bounded confidence model, opinion (O) is treated as a continuous variable ranging between the two extremes of the x-axis, which, for simplicity and without loss of generality, are set to 0 and 1. We assign to each agent a confidence bound that determines the maximum distance along the opinion axis within which they react to messages from an emitter. Emotional arousal (EA), a continuous variable, also ranging from 0 to 1 and represented on the y-axis, models the emotional levelel attached to the opinion and governs the width of each agent's confidence bound through the following relation:
\begin{equation}
d_i = 2d \cdot (1-EA_i),
\label{Eq:confidence_bound}
\end{equation}
where $d$ is the basal confidence bound of the model that represents the confidence bound of an agent with $EA = 0$. The confidence interval is considered open, hence an agent with $EA = 1$ behaves as a zealot \cite{mobilia2007role} and never reacts to a message, even if the emitter's opinion is exactly the same. Each agent, as an IFO, is assigned an initial internal phase $\phi_{i}$, uniformly distributed at random between 0 and 1, except for the pairwise and the synchronized limit cases (see Table \ref{Tab:cases}). The phases increase uniformly with a period $T$ until reaching a maximum value of 1, at which point a firing (communication) event occurs, and the phase is reset to zero. During a firing event, an agent influences other agents within its confidence area (see Fig. \ref{Fig:bands}-a) multiplying their phases by a factor $(1+\epsilon)$. This mechanism induces local synchronization, which can lead to multiple agents emitting messages simultaneously, modelling a cascade of messages \cite{piedrahita2013modeling}. Once a pair of agents achieve the same phase, they remain synchronized until the end of the simulation. Following phase updating and firing events, all agents move towards the barycenter formed by the $n$ emitters located within their confidence area, following the equation below:
\begin{eqnarray}
O_{j}(t+1) &=& O_{j}(t) + min(\alpha/d_{C_{j}}, 1) \frac{\sum_{i = 1}^{n}(O_i - O_j)}{n+1}, \\
EA_{j}(t+1) &=& EA_{j}(t) + min(\alpha/d_{C_{j}}, 1) \frac{\sum_{i = 1}^{n}(EA_i - EA_j)}{n+1},  \nonumber
\label{Eq:movement}
\end{eqnarray}
\begin{table*}
\caption{\label{Tab:cases}Table to summarize the variants of the model.}
\begin{center}
\renewcommand\arraystretch{1.8}
\begin{tabular}{| >{\raggedright\arraybackslash}p{0.2\linewidth} | >{\raggedright\arraybackslash}p{0.2\linewidth} | >{\raggedright\arraybackslash}p{0.45\linewidth} |}
\hline
\textbf{Variants of the model} & \textbf{Internal phases} & \textbf{Time step}\\  
\hline
{\bf{A.}} Pairwise
&
Agents have no internal phases
&
Two agents, one emitter and one receiver, are randomly selected. The receiver moves towards the emitter if they are within the receiver's interaction area.
\\\hline
{\bf{B.}} Sequential
&
Initial random phases $\phi_{i} \in [0, 1]$.
&
$\epsilon = 0$ Sequential firing in a consistent decreasing phase order.
\\\hline
{\bf{C.}} Broadcasted with local synchronization
&
Initial random phases  $\phi_{i} \in [0, 1]$.
&
$\epsilon \ne 0$ Firing in an order dictated by decreasing phases. Phases increase upon message reception causing message cascades and leading to progressive cluster synchronization.
\\\hline
{\bf{D.}} Fully synchronized
&
Initial phases $\phi_{i} = 1$ $\forall i$.
&
All agents firing at once since the beginning.
\\\hline

\end{tabular}
\end{center}
\end{table*}

where $d_{C_{j}}$ is the distance between the $j^{th}$ agent and the barycenter formed by the $n$ emitters inside their confidence area. The system evolves until all the interacting agents collapse into single points in the O-EA  plane, representing opinion groups. At the end of the simulation, all agents belonging to the same opinion group fire synchronously for $\epsilon > 0$. We define this firing cycle as our time unit, referred to as a stroboscopic cycle. In the case of pairwise interaction, where there is no reference agent, we define the unit time as the number of firing events divided by the system size, enabling comparison with other scenarios. See Appendix \ref{App:algorithm} for more details about the dynamics implementation.\\

In line with previous studies on the bounded confidence model, our primary focus is to investigate the formation of opinion groups, specifically assessing the number of them obtained at the end of the simulation, with a particular interest in consensus and bipartisanship, i.e. two groups. We define an opinion group as a subgroup of agents that are connected by interactions, where an agent $i$ is considered to interact with or be linked to agent $j$ if agent $i$ moves when agent $j$ emits a message, or if they are overlapped. Due to the variation in confidence interval widths resulting from different levels of emotional arousal, the resulting interaction network is directed, as shown in Fig. \ref{Fig:bands}. Therefore, we define an opinion group as a weakly connected component of this directed interaction network.

In certain cases, small groups that separate from the main clusters \footnote{While it is preferable to use therm \textit{opinion group} to refer to a group of people that share a point of view, and the therm \textit{cluster} to refer to agents with the same value of the variable opinion in the model, in this work we use the two therms interchangeably} may appear for certain parameter values. These small groups, referred to as "wings" in \cite{Deffuant_2000}, are excluded from the statistical analysis when counting the final number of opinion clusters, unless otherwise specified. To be included in the counting, an opinion group must be larger than 5\% of the size of the largest cluster.

It is worth noting that in our model, opinion groups can break but they cannot merge, in line with groups dynamics in the one-dimensional bounded confidence model \cite{gomez2012bounded}. All interactions are attractive, but by definition, two separated opinion clusters cannot interact and, as a result, they cannot attract each other to merge. Any agent that could potentially mediate the interaction between the opinion groups would necessarily be connected to both groups, resulting in a single weakly connected component instead of two. If we consider strongly connected components \cite{barandiaran2020defining}, as we do in Section \ref{Subsec:EA}, they may potentially merge.

Whenever the final outcome is not consensus, we pay close attention to the polarization of the system. The measure of polarization used in this study follows the definition provided by Esteban et al. in \cite{Esteban_1994}:
\begin{equation}
P(\pi, O) = K \sum_{i = 1}^{n}\sum_{j = 1}^{n} \pi_{i}^{1+ \beta} \pi_{j} | O_i - O_j | ,
\label{Eq:polarization}
\end{equation}
where the summation is carried out over the $n$ final opinion groups. Here, $K$ represents a normalization constant that is calculated based on the most polarized situation possible, which corresponds to two clusters of size $N/2$ each, located at the extremes of the opinion segment. The exponent $\beta$ is numerically determined in \cite{Esteban_1994} and has a value of 1.6, therefore $K = N (N/2)^{2.6}$

\section{Results and Discussion} \label{Sec:Results}
\subsection{Trajectories}\label{Subsec:traj}

\begin{figure}[h!]
\centering
\includegraphics[width=1.0\linewidth]{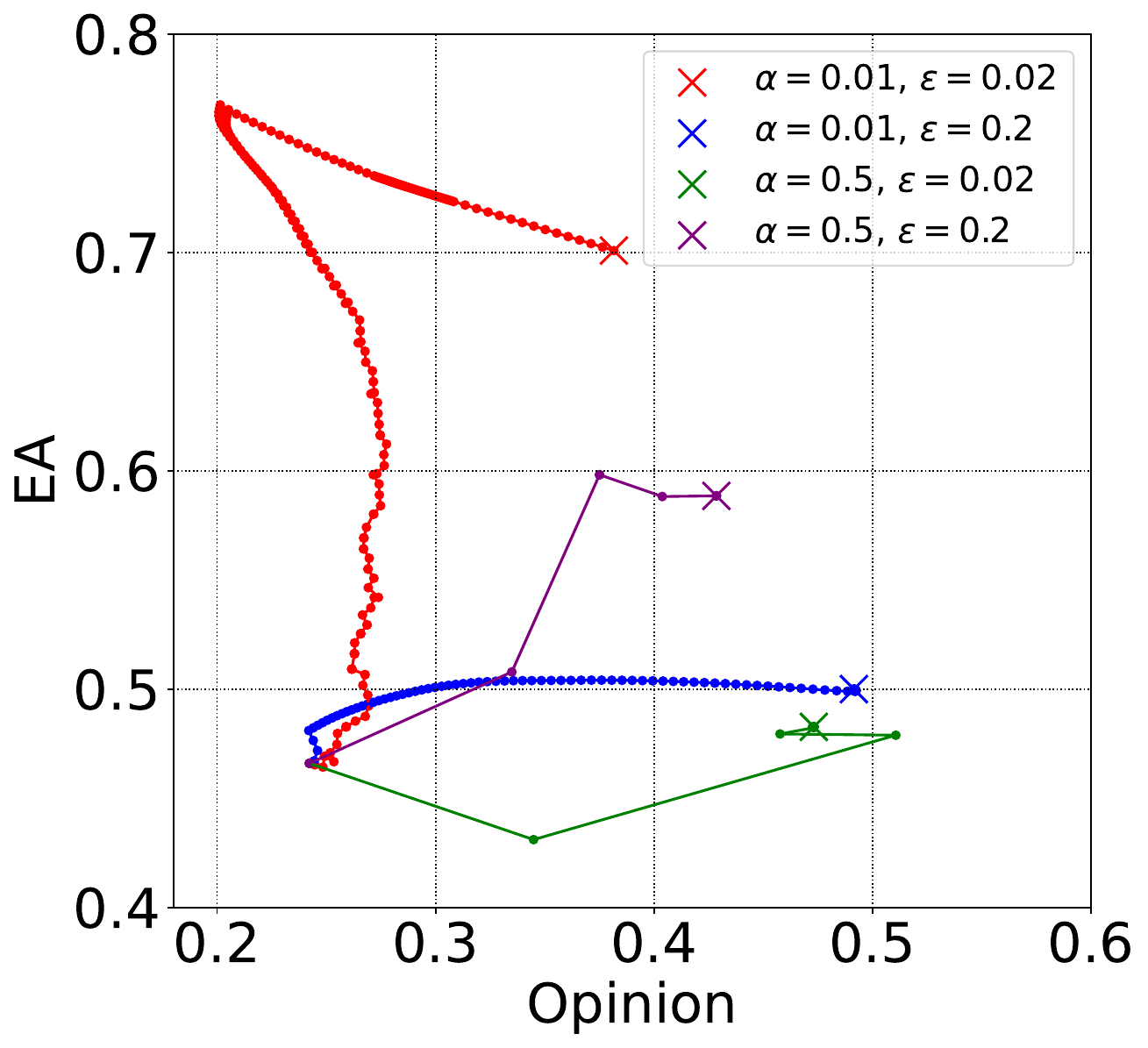}
\caption{Examples of one-agent trajectories for different step lengths $\alpha$ and synchronization factors $\epsilon$ in the plane formed by the opinion axis (vertical) and the emotional arousal axis (horizontal) for systems consisting of $N = 100$ agents with basal confidence bound $d = 0.5$.}
\label{Fig:one_traj}       
\end{figure}

%
%
We aim to investigate the impact of the different control parameters on the number of opinion clusters that emerge in the stationary state. Due to the model's complex behavior and the non-trivial interplay between these parameters, our objective is to demonstrate how the step length $\alpha$, and the synchronization factor $\epsilon$, influence the system's behavior in a particular simulation that leads to consensus. To ensure consensus is reached, we set a sufficiently large value for the basal confidence interval, $d = 0.8$, and we assign random initial positions to $N = 200$ agents, verifying that consensus is achieved in all explored cases. Under these conditions, we track the trajectory of a reference agent for four different combinations of $\alpha$ and $\epsilon$, with both small and large values for each parameter (see Fig. \ref{Fig:one_traj}). Each dot represents a displacement associated with a firing event, including any resulting cascades. Our observations reveal distinct trajectory lines and equilibrium points at which the system reaches consensus.

The red line, corresponding to low values of both $\alpha$ and $\epsilon$, exhibits an almost vertical displacement in the initial movements. The red agent increases their emotional arousal, and after a while, they simultaneously start shifting their opinion toward the left extreme of the opinion space. When the reference agent reaches the highest point in their trajectory, they suddenly changes the movement direction for both axes, approaching directly to the consensus point at $O = 0.43$ and $EA = 0.69$. The movement of the red agent can be explained by the fact that dense regions attract agents from less populated areas \cite{Hegselmann_2002}. In fact, the red agent does not move directly upwards, but rather in a random fashion during a few initial displacements, since the density of agents is still homogeneous. It is only after these initial firing events, when the average emotional arousal ($EA$) has increased due to the asymmetry in connections that favors the reception of messages from agents in the upper part of the opinion space, that the movement becomes steadily upward. The low synchronization factor contributes to the vertical movement since the emission is almost sequential at the beginning, therefore the red agent cannot potentially receive several simultaneous messages from the bulk of central agents that would pull them towards the center, as is the case when $\epsilon$ is larger (blue agent). The drift towards the left is attributed to a small group of agents located in the upper left corner of the opinion space, acting as zealots. These zealots have such a narrow interaction band that they do not receive messages from the rest of the agents. However, once the reference agent's band becomes narrow enough to stop receiving messages from the zealots, they start moving toward the rest of the group and eventually collapse at the consensus point.  The group of zealots in the upper left corner ends up separated from the large consensus group, but they represent a very small percentage of the population and are classified as a wing.\\

The blue line, corresponding to small $\alpha$ and large $\epsilon$, exhibits a behavior similar to the red line in the first few movements when the synchronization factor has not yet significantly affected the system. However, in these initial movements, synchronization allows the system to pull the zealots who formed the wing in the previous scenario and connect them to the larger group. Consequently, the consensus is achieved without wings, resulting in a reduction in the equilibration time, lower final emotional arousal, and a final opinion closer to the center at $O = 0.5$. \\

Trajectories for a large value of the step parameter $\alpha$ are represented by the green and purple lines, corresponding to a small and a large synchronization factor $\epsilon$, respectively. Both exhibit similar behavior and achieve consensus near the center of the opinion space in just a few iterations. Moreover, no wings were observed in these simulations. The green trajectory shows more frequent changes in direction due to the sequential emission characteristic of low $\epsilon$ values. In contrast, the purple line represents a scenario where the reference agent is simultaneously influenced by multiple synchronized emitters. When multiple emitters collectively attract receivers to their barycenter instead of pulling from different points in each firing event, the resulting overall displacement appears smoother, having fewer changes of direction.\\

It is important to highlight that all outcomes shown in Fig. \ref{Fig:one_traj} are significantly influenced by the initial conditions, as the position of the first emitter can dramatically affect the early dynamics, particularly for large values of $\alpha$. This strong dependence on the early dynamics has also been discussed elsewhere \cite{Kan_2022}.\\

\begin{figure}[h]
\centering
\includegraphics[width=1.0\linewidth]{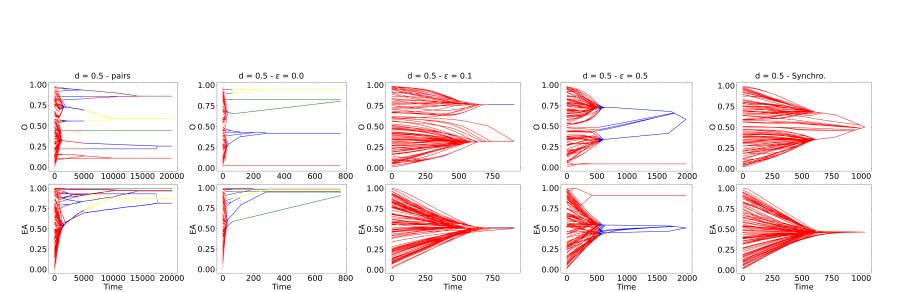}
\caption{Examples of agents trajectories for different parameters in the opinion axis (up) and the emotional arousal axis (down) for systems formed by $N = 100$ agents, $d = 0.5$ and $\alpha = 0.001$. Colors are tags denoting the belonging to an opinion cluster. We use the sequence of colors [red, blue, green, yellow, purple] to represent agents belonging to different opinion groups. All simulations start with a single connected component colored in red. When a second cluster emerges, it is designated as blue, and so forth. Agents receive color assignments at each time step, regardless of their previous color. The grey vertical dashed lines serve as visual guides to aid in identifying the time steps at which the groups fragment.
}
\label{Fig:traj}       
\end{figure}

%
%
%
In Fig. \ref{Fig:traj}, we present specific examples that illustrate the effect of synchronization on the entire system. We choose a basal bounded confidence interval of $d = 0.5$ for a system of $N = 100$ agents and set random initial positions and phases. To facilitate visualization of the agents' trajectories, we use a smaller value for the step longitude, $\alpha = 0.001$, compared to the rest of the sections. However, it is important to note that the precise value of $\alpha$ is not crucial in this section, as our aim is to provide qualitative insights into the system's behavior.\\

The first row in Fig. \ref{Fig:traj} shows the evolution of opinions as a function of time steps, while the second row depicts the corresponding emotional arousal. The colors represent the membership to an opinion cluster (weakly connected component), and they change over time as the clusters break. It is worth noting that the chosen basal bounded confidence interval is much larger than the value corresponding to the percolation transition, $d_{\text{perc}} \approx 0.02$ (see Eq. \ref{eq:percolation}). Consequently, there is a single opinion cluster (red) in all cases at $time = 0$.\\

\begin{equation}
d(N-1)-(1-d)^N\sim 1
\label{eq:percolation}
\end{equation}

The first column corresponds to the pairwise case, which is similar to the model proposed by Deffuant et al. \cite{Deffuant_2000}, with the addition of the emotional dimension. In the stationary state, the system is fragmented into five clusters, evenly distributed along the opinion axis due to the symmetric interaction along this axis. Since the interaction is also attractive the extremes of the segment $\text{O} = 0$ and $\text{O} = 1$ are devoid of agents. Conversely, along the $\text{EA}$ axis, all clusters accumulate in the upper part, with $EA > 0.75$. Specifically, the green-colored cluster, which concludes the simulation has $EA = 1$, and an opinion close to $0.5$. The average initial width of the interaction bands is approximately $0.25$, which corresponds to the consensus transition in the original model by Deffuant et al. \cite{Deffuant_2000}. However, since agents with higher levels of EA receive fewer messages, the overall system's $EA$ increases over time, reducing this average width and leading to a more fragmented outcome. The convergence time is an order of magnitude larger than that observed in any of the broadcast cases shown in the other columns.\\

The second column corresponds to the purely sequential case, with broadcasted interactions and a coupling factor  $\epsilon = 0$. This case is similar to the previous one, but the number of final clusters is reduced to four instead of five because, thanks to the multi-body interactions, agents can group more easily. Furthermore, convergence is much faster, requiring only $800$ stroboscopic cycles compared to 20.000 cycles in the pairwise case. Another notable difference is that with multi-body interactions, the system is unable to merge already separated clusters, whereas in the pairwise case, opinion groups can both fragment and reunite. Both in the sequential and the pairwise case fragmentation occurs during the early steps of the simulation.\\

The third column exhibits significant differences compared to the previous two. In the final scenario, bipartisanship is observed, with two large clusters separated by a distance close to $d = 0.5$ between them, and positioned at $d/2 = 0.25$ from the extremes of the segment, as in the original model by Deffuant et al. \cite{Deffuant_2000},. However, since the EA of both clusters is approximately 0.5, their interaction bands at that position are half the width of those in the original model. The transient regime also differs in this case, as the displacements are smoother, but the convergence time is similar to the sequential case, indicating that synchronization does not significantly affect this quantity.\\

The fourth column corresponds to a high value of $\epsilon$, resulting in rapid synchronization of agents' emissions. The outcome can be considered as a consensus, with the second cluster, consisting of only 3 agents, classified as a wing. This wing exhibits high EA and an "extreme" opinion value, close to zero. For more details on the number and position of wings depending on the model parameters, refer to Appendix \ref{App:wings}. The main opinion cluster is located at the center of the opinion space, coinciding with the equilibrium point for the consensus outcome. The last column presents a limiting case in which all agents fire simultaneously from the beginning of the simulation, and ends with a consensus with no wings. Both the convergence time and the smoothness of the trajectories are similar to the two previous examples with $\epsilon > 0$.\\

From these examples, as we witness a gradual decrease in the number of final clusters with heightened synchronization, we can infer that both multi-body interactions via broadcasted messages and synchronized emission play important roles in preventing opinion fragmentation. Furthermore, synchronization also helps in avoiding an increase in emotional arousal. 

Note that in the pairwise case, stroboscopic time defined as number of attempted interactions/firing events over $N$.
\onecolumngrid

\twocolumngrid

In all cases, we can distinguish between two movement regimes: the first, characterized by numerous displacements at each time step, and the second, in which most agents have already collapsed into a few opinion clusters that evolve slowly until reaching the stationary state.

\subsection{Number of clusters}\label{Subsec:clust}

Here, we investigate the number of opinion groups (after filtering out the wings) for various model parameter values. We perform 300 independent simulations with different initial positions and phases for each parameter set and present the results in Figure \ref{Fig:swap_d}, which illustrates the distribution of different outcomes. We focus on four distinct stationary states: consensus, bipartisanship, and fragmentation into three or four opinion groups. As expected, the average number of final opinion groups increases when decreasing the parameter $d$ across all parameter combinations. However, the specific transition to consensus or the value of $d$ at which a particular outcome has the highest probability of occurring depends on the step size, synchronization factor, and system size.
\subsubsection{Variation on $d$ for different fixed values of the remaining parameters} \label{Subsubsec:d}   
We observe that increasing $\alpha$ leads to broader and noisier distributions, making it impossible to ensure consensus in any case for $\alpha = 0.5$. These effects can be attributed to a higher sensitivity to initial conditions, which is related to the fact that agents experience larger movements when they interact. Consequently, the system tends to break into smaller clusters more easily during the initial steps of the simulation. 

\begin{figure}[h!]
\centering
\includegraphics[width=1.0\linewidth]{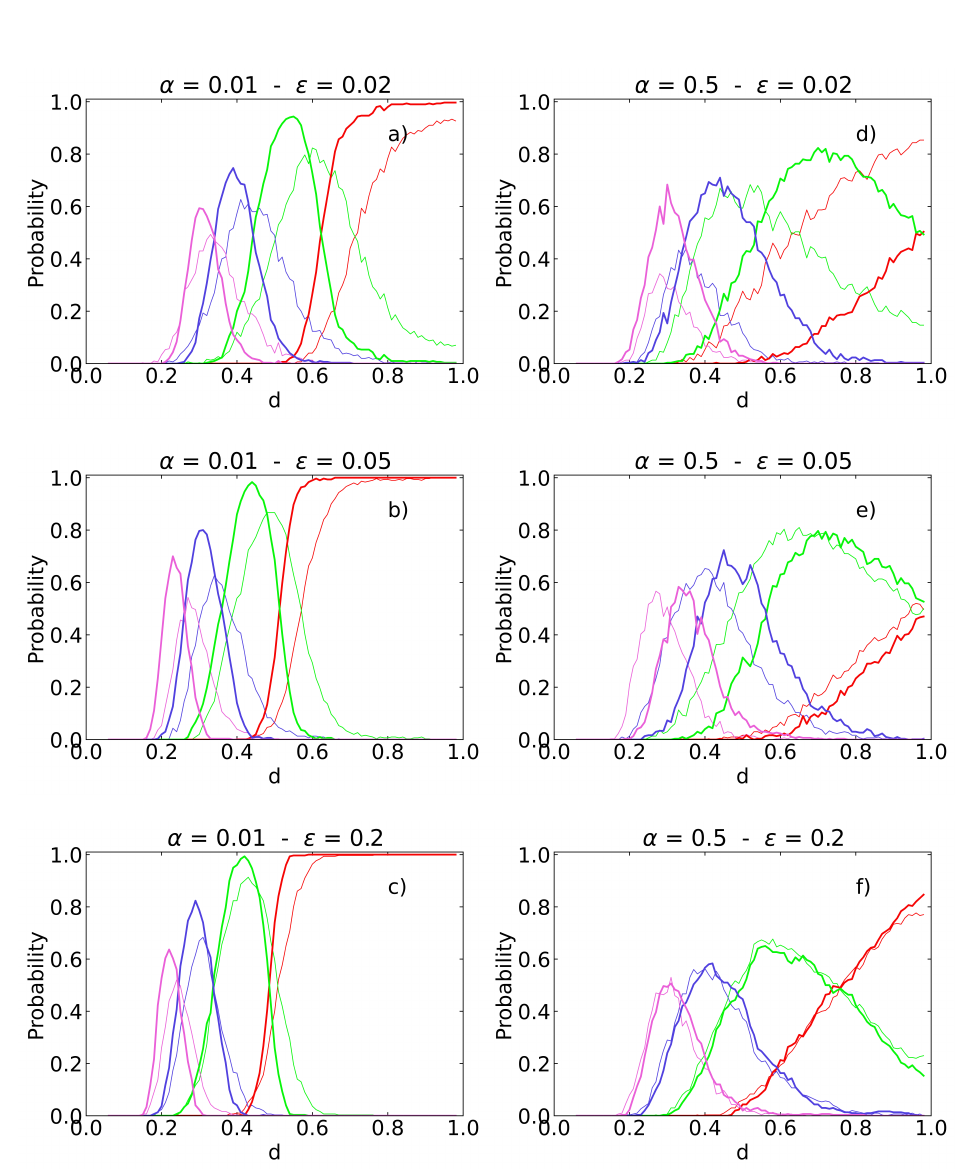}
\caption{Proportion of final scenarios as a function of the parameters. Colors denote the different outcomes: red for consensus, green for bipartisanship, blue for fragmentation into three opinion groups, and pink for fragmentation into four opinion groups. Results obtained for a system of $N = 200$ agents (thin line) and $N = 1000$ agents (thick line), running 300 simulations with different initial conditions for each parameter set.}
\label{Fig:swap_d}       
\end{figure}

The system size also plays a significant role in the outcome distributions. For $\alpha = 0.01$, increasing the system size $N$ decreases the value of the basal confidence bound $d$ required for the consensus transition, and all the peaks corresponding to fragmented outcomes shift to the left. Surprisingly, for $\alpha = 0.5$ and low $\epsilon$, we observe the opposite effect, and for $\alpha = 0.5$ and $\epsilon = 0.2$, there appears to be no noticeable effect of the system size. It is worth noting that having a sharper transition to consensus for smaller $N$, as depicted in Fig. \ref{Fig:swap_d} - (d), is not a commonly observed result. The underlying reason is that a higher agent density amplifies fragmentation caused by sequential firing during the initial steps of the simulation, as a larger number of receivers are affected by these initial interactions. From a social perspective, this implies that denser initial groups are more likely to become fragmented if individuals strongly influence one another and several agents' opinions are significantly altered after a single interaction, without considering alternative viewpoints. However, the study of other values of $N$ to determine the scaling behavior is out of the scope of this paper. 

On the other hand, increasing $\epsilon$ has a smoothing effect on the system, closing the gap between the $N = 200$ and $N = 1000$ curves and reducing the noise. In terms of the consensus transition, we once again observe differences depending on $\alpha$. For $\alpha = 0.01$, increasing $\epsilon$ subtly enhances the position of the consensus transition, and shifts to the left the probability peaks corresponding to fragmented outcomes. However, for $\alpha = 0.5$, the effect of $\epsilon$ depends on $N$, with a more pronounced effect observed in smaller systems. Notably, the transition to consensus occurs at larger $d$ for $\epsilon = 0.05$, deviating from the monotonic trend observed in other cases.\\

The interplay of the model parameters becomes non-trivial, particularly when the system undergoes larger changes in each interaction (higher step length $\alpha$) and the movement does not involve multiple emitters firing simultaneously (lower coupling factor $\epsilon$). In such cases, the dynamics of the system exhibit a complicated behavior and is more sensitive to variations in the initial conditions, leading to a larger variety of outcomes.\\

\subsubsection{The synchronized case}\label{Subsubsec:sync}  
\begin{figure}[h!]
\centering
\includegraphics[width=1.0\linewidth]{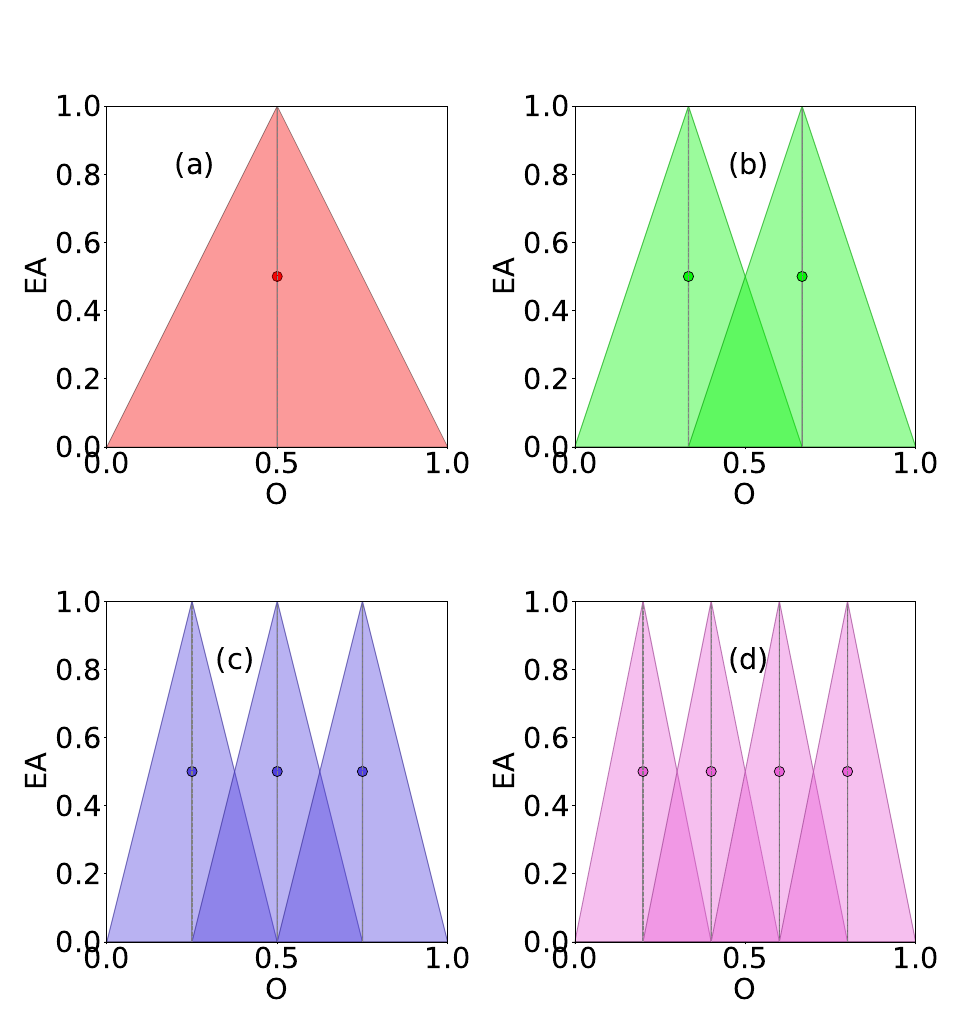}
\caption{Basins of attractions and theoretical equilibrium points of (a) consensus, (b) bipartisanship, (c) fragmentation in 3 clusters, and (d) fragmentation in 4 clusters in the synchronized case.}
\label{Fig:basins}       
\end{figure}

Here we present the simulation results for the synchronized limit case where all agents emit messages at once at every time step since the beginning. Since this variant lacks random initial phases and depends only on the initial positions, we can provide a semi-analytical estimation for the proportion of clusters based on the following assumptions (see Fig. \ref{Fig:basins}):
\begin{itemize}
\item For each outcome, there exists an equilibrium position in the opinion axis for the opinion clusters, given by the original model, which corresponds to $\bar{O_i} = i/(n+1)$, where $n$ is the number of final clusters.
\item Each equilibrium position $\bar{O_i}$ has an associated area $2d\cdot L/2 = d$ that corresponds to the affecting triangle of an agent placed at this equilibrium point ($L = 1$ corresponds to the height of the opinion space).
\item The normalized total attraction area associated with all the equilibrium points of a given outcome is $A(n, d) = d/D_n$, where $D_n = 2/n$ is twice the distance between the opinion axis boundary and the first equilibrium point, representing the maximum base of a triangle that contains only one equilibrium position.
\item The probability of a given outcome is equal to the probability that initially at least half of the agents lie within the attraction area associated with this outcome, minus the probability that initially at least half of the agents lie within the attraction area associated with an outcome with fewer clusters. Since initial positions are sampled from a uniform distribution, we have:
\end{itemize}
\begin{widetext}
\begin{equation}
P \left( n \right) = \sum_{k = N/2}^{N} A \left(n, d \right)^{k}  \left(1 - A \left(n, d\right)\right)^{N-k} - \sum_{\eta = 1}^{n-1} \left( \sum_{k = N/2}^{N} \left(A\left(\eta, d \right) \right)^{k}  \left(1 - A \left(\eta, d \right) \right)^{N-k} \right) \; ; n \ge 1
\label{Eq:sync}
\end{equation}
\end{widetext}

\begin{figure}[h!]
\centering
\includegraphics[width=1.0\linewidth]{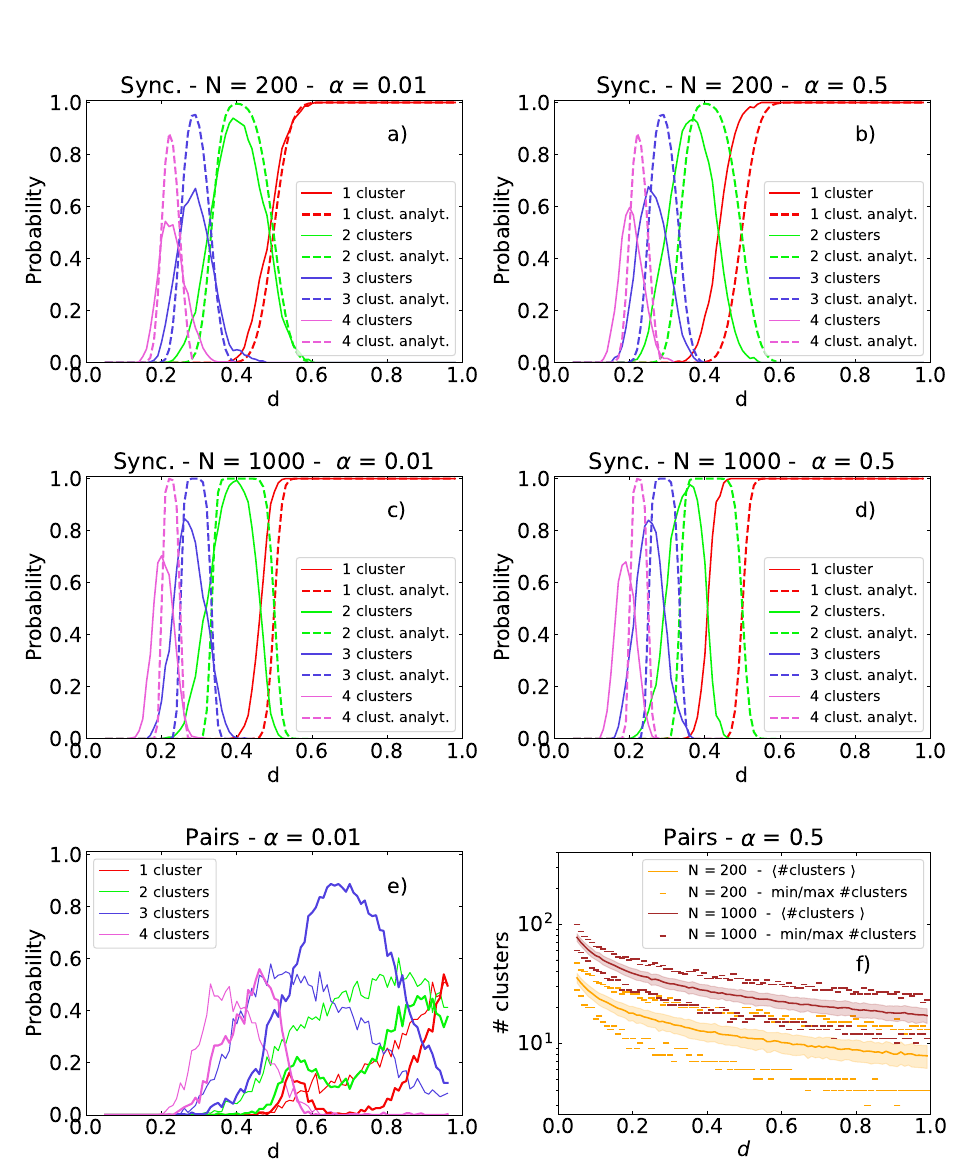}
\caption{Proportion of final scenarios as a function of the basal confidence bound $d$ for a) synchronized interactions, step longitude $\alpha = 0.01$ and system size $N = 200$, (b) synchronized interactions, $\alpha = 0.5$ and $N = 200$, c) synchronized interactions, $\alpha = 0.01$ and $N = 1000$, d) step $\alpha = 0.5$ and  $N = 1000$ and e) pairwise interactions, $\alpha = 0.01$ and $N = 200$ (thin line), $1000$ (thick line). f) Average maximum and minimum number of final clusters as a function of $d$ for pairwise interactions, $\alpha = 0.01$ and $N = 200, 1000$. Results were obtained for 300 simulations with different initial conditions.}
\label{Fig:clust_sync_pairs}     
\end{figure}

Based on these assumptions, the probability of achieving consensus is 1 if at least $N/2$ agents lie initially inside the affectation triangle whose peak is at $(\text{O, EA} )= (L/2, 1)$, and has a base length $L = 1$ (Fig:basins)-a)). This estimation tends to overestimate the value of $d$ for the transition to consensus and the position of the peaks for the other outcomes, as observed in our simulations for all parameter values explored. However, it provides a closer approximation for lower step longitude values ($\alpha$) as shown in Fig. \ref{Fig:clust_sync_pairs}. It is important to note that this approximation is not influenced by the presence of wings (which are more abundant for higher $\alpha$) since not filtering the wings from the results obtained with the simulations only introduces noise and distorts the results. But with the wing filtered the probability distributions exhibit no noise, even for $\alpha = 0.5$ (Fig. \ref{Fig:clust_sync_pairs}-right column), indicating that the sensitivity to initial conditions is primarily related to the phases that determine the emitter in the first time steps. With perfect synchronization from the beginning, a larger step does not promote system fragmentation but drives it toward consensus at smaller values of the basal confidence interval $d$.\\

If we modify the lower bound of the sum in Eq. \ref{Eq:sync}, requiring a greater number of agents to lie initially within the considered attraction area, we can improve the estimation. Although we see that the analytical expression for this bound is an increasing function of $\alpha$ and $N$, the exact formula is beyond the scope of this paper, however, some approximated corrections to the original estimation can be found in Appendix \ref{App:sync}. What occurs is that, while agents may not be initially within this area, the attractive dynamics lead them to reach it at earlier times for larger $\alpha$. Since they do so in a cohesive manner, influenced by all other agents simultaneously, this does not result in fragmentation. The analytical approximation captures the shape of the transition, which becomes sharper for larger $N$ but remains centered around the same $d$ values. Furthermore, it accurately represents the tendency for the peaks of other outcomes to decrease and narrow.
\subsubsection{The pairwise case} \label{Subsubsec:pairs}    
%
To establish a connection between our results and the classical case involving pairwise interactions, we conducted additional investigations on our two-dimensional opinion space with pairwise interactions. Since the pairs can be chosen in any order, the final outcome can be dramatically affected by the particular pair selection. For $\alpha = 0.01$ we observe consensus for certain values of $d$ when $N = 200$ (thin lines in Fig. \ref{Fig:clust_sync_pairs}-e), but it cannot be guaranteed for any $d \in [0.05, 1]$, resembling the behavior exhibited for $\alpha = 0.5$ and $\epsilon = 0.05$ but with broader peaks. However, a non-monotonic behavior in the proportion of consensus and bipartisanship, unique to this limit case and parameter combination, emerges for $N = 1000$ (thick lines). In both cases, two peaks are observed, with the peak for three opinion clusters being the highest. This characteristic, with an approximate probability of 90\% for having 3 opinion clusters in the stationary state when $d \approx 0.7$, is also exclusive to this specific parameter combination.\\

For $\alpha = 0.5$, the final configuration becomes significantly more fragmented. The number of opinion groups, which increases with higher $N$, decreases supra-exponentially as $d$ varies within the range $d = [0.05, 1]$ (see Fig. \ref{Fig:clust_sync_pairs}-f). The maximum average number of clusters, corresponding to $d = 0.05$, is approximately $\langle n \rangle = 40$ for $N = 200$ and $\langle n \rangle = 80$ for $N = 1000$. This proportion is maintained for all $d$ values, reaching a minimum around $\langle n \rangle = 10$ for $N = 200$ and $\langle n \rangle = 20$ for $N = 1000$ at $d = 1.0$.\\

A higher overall density implies having more agents in the extremes on the opinion axis, which could lead to the formation of 3 clusters more easily than when $N$ is smaller.

The behavior of the pairwise variant, which exhibits significantly more noise, is attributed to the wide range of confidence intervals associated with randomly selected interacting pairs. We explain the formation of double peaks for consensus and bipartisanship when $N = 1000$ in Subsec. \ref{heatmap_alpha-eps}.

The most important changes in this limit case are due to the value of $\alpha$. The system size also plays a qualitatively significant role in the basins of attraction for $\alpha = 0.01$.

\subsubsection{Variation on $\epsilon$ for different fixed values of the rest of parameters} \label{Subsubsec:eps}
\begin{figure}[h!]
\centering
\includegraphics[width=1.0\linewidth]{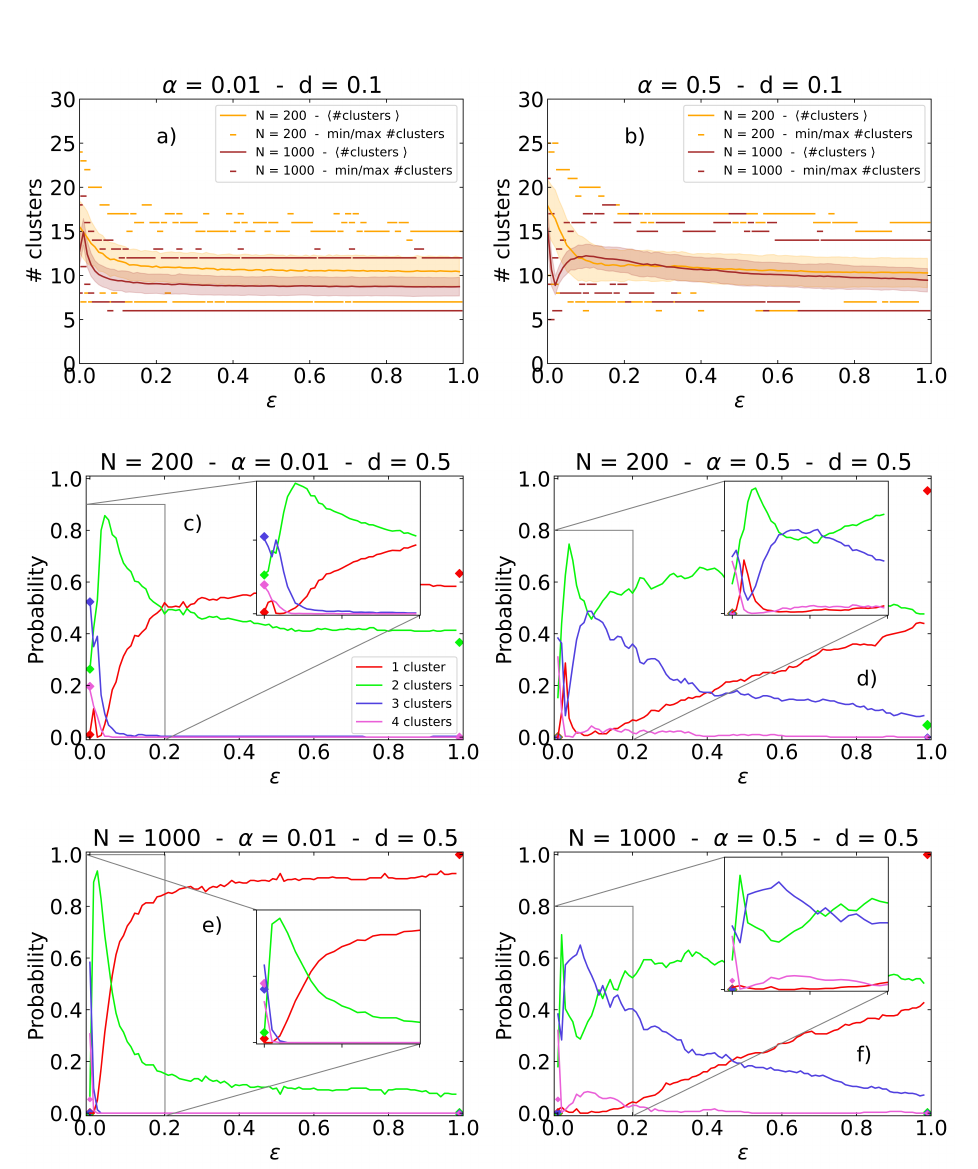}
\caption{Average, maximum, and minimum number of clusters as a function of the coupling factor $\epsilon$ for different parameters values. Results obtained by performing 300 simulations with different initial conditions for each parameter set, except for the inset plots in panels c) and d), where 1000 simulations were performed with finer resolution in parameter values.}
\label{Fig:swap_eps}       
\end{figure}

Here we conduct a detailed investigation on the impact of the coupling factor $\epsilon$ within the range $[0, 1]$ (including the sequential limit case $\epsilon = 0$) for two values of the maximum confidence interval, $d = 0.1$ and $d = 0.5$ (see Fig. \ref{Fig:swap_eps}). We compare the results of the sequential limit case at $\epsilon = 0$ with those obtained for the pairwise limit case, as well as the results for $\epsilon = 1$ with the synchronized limit case. The pairwise and synchronized cases are represented by diamond dots at $\epsilon = 0$ and $\epsilon = 1$, respectively. It is important to note that each value of the parameter $d$ corresponds to a distinct region of potential outcomes. In the case of $d = 0.5$, we separate the results obtained for each system size.

%
For $d = 0.1$, we do not expect to observe any of the four focused outcomes. Therefore, we pay attention to the average number of opinion groups, its standard deviation, and the maximum and minimum number of clusters for each value of $\epsilon$. We clearly distinguish two regimes for both values of $\alpha$: a coupling-dependent region for $\epsilon$ below approximately 0.1, and above this value, a region that is nearly independent of the coupling and exhibits an average number of opinion groups around $n = 10 \pm 2$ for both system sizes and step longitudes. The coupling-dependent regime is more extensive for $\alpha = 0.5$ and depends on the system size. Specifically, for $N = 1000$ and $\alpha = 0.01$, the average number of opinion groups peaks at $\epsilon \approx 0.02$ and shows a relative minimum around the same value for $\alpha = 0.5$. On the other hand, for $N = 200$, there is a monotonic decay for both values of $\alpha$. When $\alpha = 0.01$, $\langle n \rangle$, $n_{\text{min}}$, and $n_{\text{max}}$ are larger for $N = 200$ than for $N = 1000$ for all $\epsilon$. However, when $\alpha = 0.5$, these quantities are more similar for each system size, and their magnitudes depend on the value of $\epsilon$. In general, in this fragmented region of parameter space, higher agent densities and larger coupling factors tend to result in less fragmentation, except for the specific values of $\alpha = 0.5$ and $\epsilon \in [0, 0.4]$, approximately.\\

%
When approaching the region where consensus is observed, at $d = 0.5$, we again observe non-monotonic behaviors for low values of the coupling factor. For $\alpha = 0.01$ (see Fig. \ref{Fig:swap_eps} (c) and (e)), the probability of consensus transitions from a value close to zero for $\epsilon \to 0$ to a finite probability, which is higher for larger system sizes. The probability of bipartisanship is complementary to the probability of consensus, as fragmentation into three or four clusters, despite having some representation at $\epsilon \to 0$, rapidly decays to zero with increasing coupling factor. The shape of the curves remains unchanged when increasing $N$, but the transition to consensus becomes sharper and the bipartisanship peak becomes narrower and higher. As a result, fragmentation into three or four opinion groups becomes more residual compared to the smaller system size. At $\epsilon = 0$, the sequential limit case yields results that closely resemble those obtained for the pairwise limit case, particularly when $N = 200$. On the other hand, when the coupling factor is 1, the results resemble those obtained for the synchronized case for both system sizes.\\

%
%
On the contrary, for $\alpha = L/2 = 0.5$, the pairwise and synchronized limit cases do not match the results for $\epsilon = 0$ and $\epsilon = 1$ (see Fig. \ref{Fig:swap_eps} d, f). The probability curves exhibit more noise due to the larger step length and show a non-monotonic behavior for low values of $\epsilon$. The probability of consensus is zero at $\epsilon = 0$ for both system sizes and increases linearly with $\epsilon$ until it reaches a value around 0.4. Consensus also exhibits a narrow peak of height 0.3 approximately at $\epsilon = 0.1 \pm 0.1$ for $N = 200$. Bipartisanship shows a peak and a minimum at $\epsilon < 0.1$ for both system sizes. Concurrent with the minimum for bipartisanship, there are peaks for the outcomes with 3 and 4 opinion groups, which are higher and occur at lower values of the coupling factor for larger $N$. However, the effect of $N$ is only noticeable for low $\epsilon$, while in the range $\epsilon > 0.2$, there are no significant differences between the two system sizes. In fact, the behavior is similar to the case with $\alpha = 0.01$, but with increased fragmentation across the entire range of studied $\epsilon$ values and a much slower transition to consensus. It is interesting to note that for this value of the maximum confidence bound $d = 0.5$, an increase in $\epsilon$ can promote fragmentation for any system size when the step length is large enough.

\subsection{Emotional Arousal and Fragmentation}\label{Subsec:EA}
%
We have generated $\epsilon$-$d$ and $\epsilon$-$\alpha$ phase diagrams to analyze the final average number of clusters and the average level of emotional arousal $\langle EA \rangle$. We observe a significant correlation between $\langle EA \rangle$ and the ratio between the final and initial average number of strongly connected components $\langle n \rangle_{fin}/\langle n \rangle_{ini}$ when wings were not filtered. However, applying the wing filtering process disrupts this correlation. The average level of emotional arousal remained relatively consistent across all parameter ranges since the wings, being small by definition, do not contribute significantly to the overall average. Instead, the ratio $\langle n \rangle_{fin}/ \langle n \rangle_{ini}$  could decrease or remain the same, depending on the number of wings filtered out (it cannot increase because wings are only defined for the final state). Furthermore, the presence of wings exhibited a strong dependence on the model parameters, as discussed in Appendix \ref{App:wings}. Note that in this context we are considering strongly instead of weakly connected components, therefore the ratio $\langle n \rangle_{fin}/ \langle n \rangle_{ini}$  can be smaller than one (i. e. groups can merge).
%
\subsubsection{Phase diagram $d$-$\epsilon$}\label{heatmap_d-eps}

\begin{figure}[h]
\centering
\includegraphics[width=\linewidth]{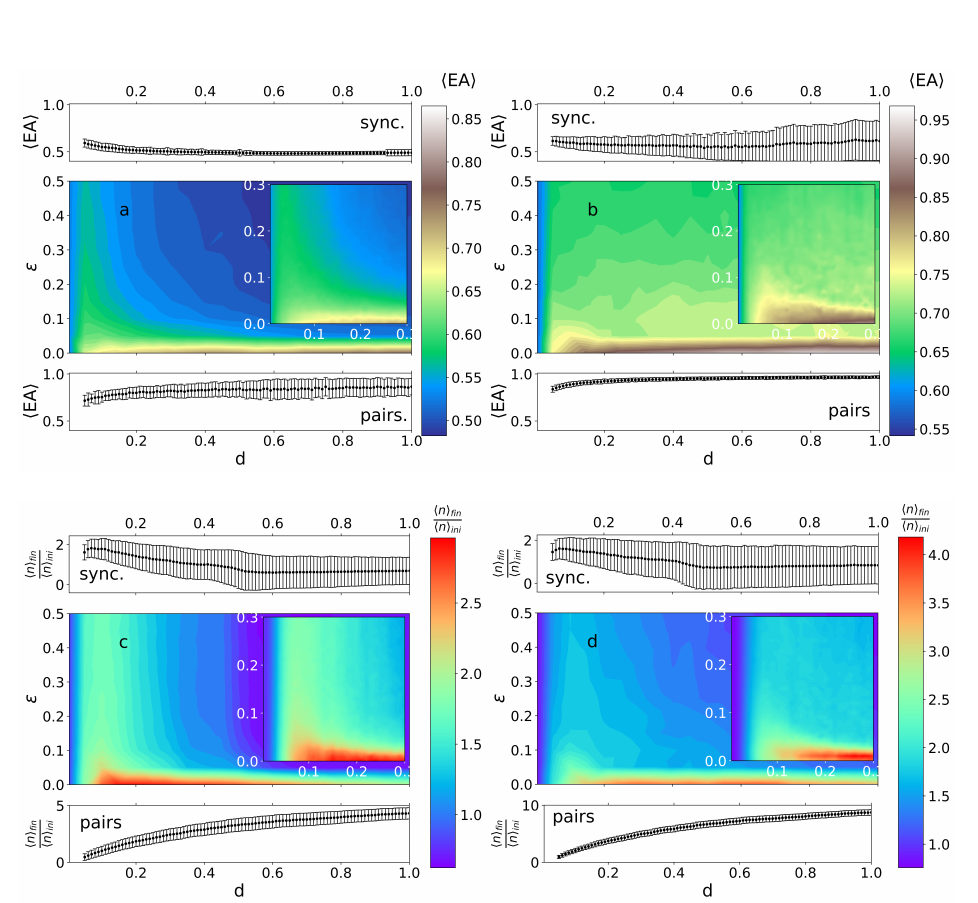}
\caption{Final values of average emotional arousal $\langle EA \rangle$ (upper row) and ratio between average number of final and initial opinion groups $\langle n \rangle_{fin}/ \langle n \rangle_{ini}$ (lower row), for a range of basal confidence bound $d$ and coupling factor $\epsilon$ when the step longitude $\alpha = 0.01$ (a, c) and $alpha = 0.5$ (b, d). Results correspond to a system formed by $N=200 $ agents and are averaged over 100 simulations with different initial conditions. The inner plot shows a smaller range with more resolution. The upper and lower subpanels in each subfigure correspond to the synchronized and pairwise cases for the corresponding set of parameters.}
\label{Fig:heatmap_d_eps}       
\end{figure}

We present the results for two values of the step longitude $\alpha = 0.01, 0.5$ in the range of $d \in [0.005, 1.0]$ and $\epsilon \in [0, 0.5]$, and compare them to those obtained for the synchronized and the pairwise cases (\ref{Fig:heatmap_d_eps}). We first analyze the phase diagrams for the average final values of the emotional arousal and the ratio between initial and final strongly connected components as a function of $d$ and $\epsilon$ (shown in Fig. \ref{Fig:heatmap_d_eps}-a and c, respectively), obtained for small $\alpha$, where we can clearly distinguish four regions:

\begin{enumerate}
\item The vertical band on the left corresponds to $d < 0.02$, marking the percolation transition (see Sec. \ref{Sec:the_model}). In this region, the dynamics have a minimal effect due to the small confidence interval, resulting in a highly fragmented situation with small opinion groups scattered across the O-EA plane. Both magnitudes, $\langle EA \rangle$, and $\langle n \rangle_{fin}$, remain unchanged from the beginning to the end of the simulation.

\item  The brown region (panel a) and red region (panel b) corresponding to $d > 0.02$ and low $\epsilon$. However, there are discrepancies in the shape of the figures for high $d$. While $\langle EA \rangle$ increases monotonically with $d$, the behavior of the ratio of strongly connected components varies.

\item  The upper-right area represents the consensus region, where $\langle EA \rangle$ and $\langle n \rangle_{fin} \le 1$. Here $\langle EA \rangle$ is approximately equal to the initial value of 0.5. The reason why $\langle n \rangle_{fin} $ can be smaller than one is that strongly connected components can merge, contrarily to the weakly connected components examined in Sec. \ref{Subsec:traj}

\item The transition region, colored in green in both panels. This zone exhibits a relative maximum along the $\epsilon$ axis at around $d \approx 0.05$, just above the percolation threshold. In this region, both $\langle EA \rangle$ and $\langle clusters \rangle_{final}$ decrease with increasing $\epsilon$ towards a saturation value that depends on $d$. This decay is faster for larger values of $d$.
\end{enumerate}

Overall, these results demonstrate the dependence of the system behavior on the values of $d$ and $\epsilon$, revealing distinct regions characterized by different dynamics and outcomes. The profiles of both $\langle EA \rangle$ and $\langle n \rangle_{fin}/ \langle n \rangle_{ini}$ are generally similar, except when we are close to the sequential case, in the limit $\epsilon \to 0$ (region II). In this limit,  $\langle EA \rangle$ monotonically increases with $d$ (as observed in the pairwise case), and $\langle n \rangle_{fin}/ \langle n \rangle_{ini}$ exhibits a relative maximum around $d = 0.3$ and then decays towards $\langle n \rangle_{fin}/ \langle n \rangle_{ini} \approx 2$. When we are close to the sequential case, an increase in $d$ primarily benefits agents located in the upper part of the O-EA plane. These agents act as zealots, influencing the rest of the agents and causing fragmentation into multiple opinion groups with high emotional arousal. Eventually, with sufficiently high $d$ fragmentation can be reduced, but the average emotional arousal will keep increasing with the confidence interval, resulting in narrower confidence bounds. Consequently, a new agent introduced into the system after reaching the stationary state would face more difficulties in being listened to by the existing agents. When the coupling factor increases the situation reverses and both  $\langle EA \rangle$ and $\langle n \rangle_{fin}$ decrease monotonically with $d$.

Since the coupling factor $\epsilon$ leads to synchronization, it is not surprising that the results tend towards the synchronized limit case for high values of $\epsilon$. As previously shown, the outcome in the synchronized case is always consensus for $d > 0.5$, therefore the fluctuations observed in this range are only caused by the different number of strongly connected components in the initial state. We observe as well an increase in fluctuations around $d = 0.5$ that corresponds to the transition to consensus. It is worth noting that highly synchronized systems tend to have average levels of emotional arousal close to 0.5, even for fragmented outcomes, whereas systems without synchronization exhibit considerably higher emotional arousal levels despite having less fragmentation.\\

Interestingly, the results for $\epsilon = 0$ tend to resemble those obtained for the pairwise case, at least in terms of $\langle EA \rangle$. However, in terms of the ratio of opinion groups, the correlation only exists for low values of $d$. While $\langle n \rangle_{fin}/\langle n \rangle_{ini}$ starts decreasing at higher $d$ for $\epsilon = 0$, it continues to increase until $d = 1.0$ in the pairwise case. Filtering the wings does not reverse this tendency but rather smoothens it (see appendix \ref{App:wings}), indicating that wings play a more significant role in the pairwise case compared to the sequential case.

%
When $\alpha = 0.5$  (Fig. \ref{Fig:heatmap_d_eps} (b) and (d) the results are resemblant but with some differences. Fragmentation and average emotional arousal are, in general, higher except below the percolation at $d = 0.02$ (region I), since agents do not interact, regardless of the step longitude. Only for large values of $d$ and $\epsilon$, where we have consensus, $\langle n \rangle_{fin}/ \langle n \rangle_{ini}$ drops to 1. The relative maximum in region III disappears in the $\langle EA \rangle$ heatmap and, above the percolation $\langle EA \rangle$ decreases depending only on $\epsilon$. The ratio of opinion groups, on the other hand, does experiment a decrease with both $d$ and $\epsilon$ when $d > 0.02$ (regions III and IV). The most noticeable difference is the overall increase in $\langle EA \rangle$, which becomes independent of $d$ above the percolation and is larger than $0.5$, even at $d = 1$ and $\epsilon = 0.5$, where consensus is guaranteed

So, the way agents with a high $EA$ drag up the other agents via their interactions 
has a direct correlation with the number of final opinion groups, taking into account the minorities (usually with a high $\langle EA \rangle$, as we can see in Appendix \ref{App:wings})
Excluding the pairwise case, the highest level of $EA$ is not reached in either the most fragmented situation or the least one, but in the parameter zone that corresponds to sequential firing and large basal confidence bound, and is always larger for larger $\alpha$.
\onecolumngrid

\twocolumngrid

\begin{figure}[h!]
\centering
\includegraphics[width=0.9\linewidth]{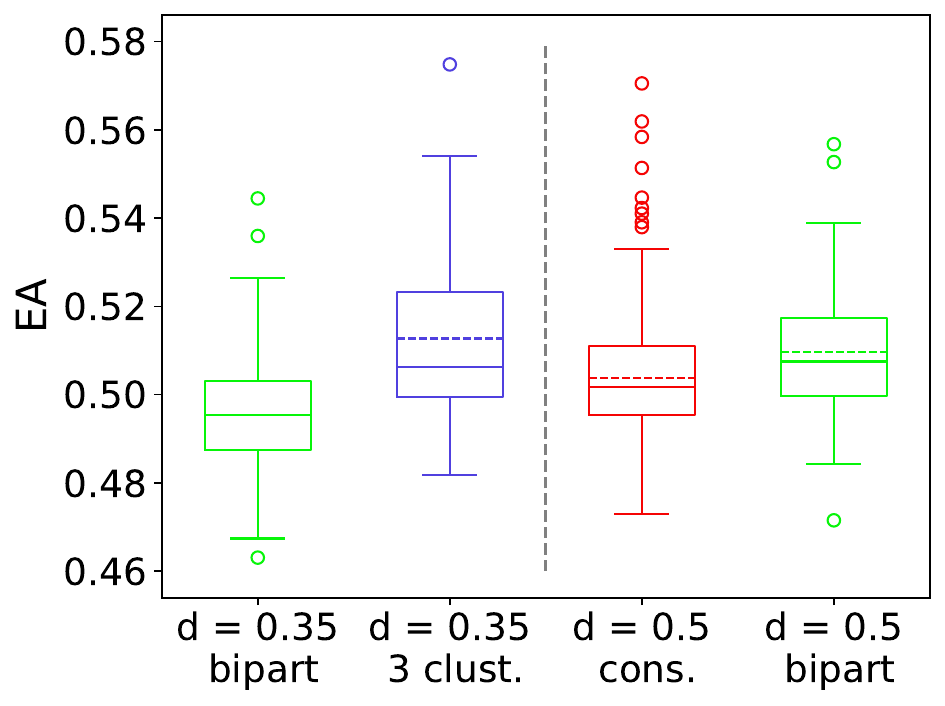}
\caption{Final EA distributions of a system of $N = 1000$ agents, longitude step $\alpha = 0.01$, synchronization factor $\epsilon = 0.2$ and different values of the basal confidence bound $d$ for those repetitions that end up in consensus, bipartisanship and 3 clusters. Results obtained for 300 simulations with different initial conditions for each value set of parameters. }
\label{Fig:boxplot}       
\end{figure}

When the set of parameters yields multiple outcomes, the most fragmented scenario corresponds to the one exhibiting a higher average emotional arousal $\langle EA \rangle$, as we see in Fig. \ref{Fig:boxplot}). This observation is drawn from examining the EA distributions for a system of $N = 1000$ with a step length of $\alpha = 0.01$, across two values of the basal confidence bound $d$. These values result in a combination of bipartisanship and fragmentation into 3 clusters ($d = 0.35$), and a mix of consensus and bipartisanship ($d = 0.5$).
%
%
\subsubsection{Phase diagram $\alpha$-$\epsilon$}\label{heatmap_alpha-eps}
We examine the range of $\alpha \in [0.005, 0.5]$, since smaller or larger step longitudes are not interesting from a social point of view, and  $\alpha >0.5$ would not be comparable with the classical bounded confidence model.

The system behaviors when varying $\alpha$ and $\epsilon$ are more straightforward, but the correlation between the final $\langle EA \rangle$ and $\langle n \rangle_{fin}/ \langle n \rangle_{ini}$ hold for all the parameter range explored. The coupling factor always diminishes $\langle EA \rangle$ making it tend to average initial value 0.5, while alpha always increases it. However, when $\alpha$ is low, decay of $\langle EA \rangle$ occurs for $\epsilon < 0.05$ and is fast, and this transition becomes smoother as $\alpha$ increases. 
\begin{figure}[h!]
\centering
\includegraphics[width=1.0\linewidth]{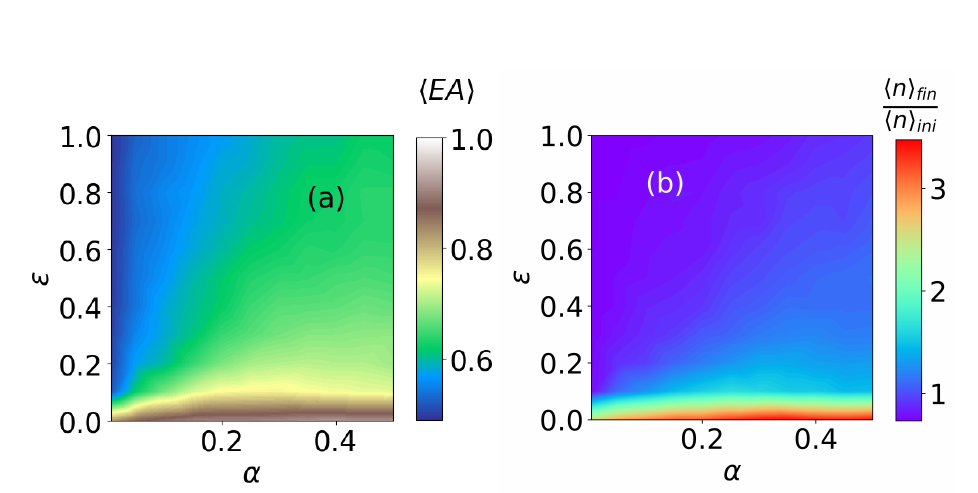}
\caption{Final values of a) average emotional arousal $\langle EA \rangle$ and b) ratio $\langle n \rangle_{fin}/ \langle n \rangle_{ini}$ for a range of the step longitude $\alpha$ and coupling factor $\epsilon$. Results correspond to a system formed by $N=200 $ agents and are averaged over 100 simulations with different initial conditions. The basal confidence bound is set to $d = 0.5$.}
\label{Fig:clust_ratio}       
\end{figure}
For $d = 0.5$ we confirm that $\langle EA \rangle$ is higher at low $\epsilon$ for any $\alpha$
%
A larger confidence bound implies a higher number of interactions, in particular with agents that have a higher EA and, since their $k_{out}/k_{in}$ degree is larger, they are privileged from a communicative point of view. Therefore a large $d$ leads to a low number of clusters since agents interact among them, regardless they are far away on the opinion axis. During the trajectory to a consensus, however, they considerably shrink their confidence interval and they lose their mediator role.

The fact that these effects are contraposed is the cause of the double peak for consensus and bipartisanship in the pairwise case for $N = 1000$ and $\alpha = 0.01$ (see Fig. \ref{Fig:clust_sync_pairs} (e). For instance, in the approximate interval $d \in (0.6, 0.8)$ agents have a basal confidence bound $d$ large enough to ascend rapidly in the O-EA. However, this parameter it is not large enough to maintain a wide confidence interval, hindering the achievement of consensus. Counterintuitively, a lower value of $d$ can result in consensus, as the upward movement is not as fast. Synchronization prevents this from occurring by diminishing the impact of zealots (i.e., agents with high EA). This is because individuals simultaneously consider not only the viewpoints of highly emotional agents but also the arguments presented by less emotional individuals.

\subsection{Polarization}\label{Subec:polarization}

Polarization reaches its theoretical peak value $P = 1$ for the bipartisan outcome when two clusters of size $N/2$ are positioned at $O = 0$ and $O = 1$ (see Eq. \ref{Eq:polarization}). In our simulations, we observe a maximum average value of $\langle P \rangle = 0.5$, achieved in the synchronized scenario for the range of $d$ where the outcome is consistently bipartisan, regardless of the initial conditions or the step length $\alpha$. In contrast, other cases exhibit a mixture of outcomes, resulting in a reduced overall average polarization.
%
%
\subsubsection{Low $\alpha$}\label{Subsubsec:polarization_low_alpha}
For smaller step lengths $\alpha$, the average polarization exhibits a distinct peak around $d = 0.4$ across all synchronization factors $\epsilon > 0.1$. This aligns with the region where the system tends to converge towards a bipartisan outcome more frequently. Notably, this peak shifts towards higher $d$ values for lower $\epsilon$, correlating with a higher tendency for fragmentation as the system approaches the sequential limit. As highlighted in Section \ref{Subsubsec:sync}, the synchronized case shows a sharp transition towards consensus, here, we observe a rapid decline in polarization at the value of the basal confidence bound $d$ where this transition occurs. On the other hand, the pairwise case exhibits more fluctuations and a smoother decrease in polarization when increasing $d$.

\begin{figure}[h!]
\centering
\includegraphics[width=1.0\linewidth]{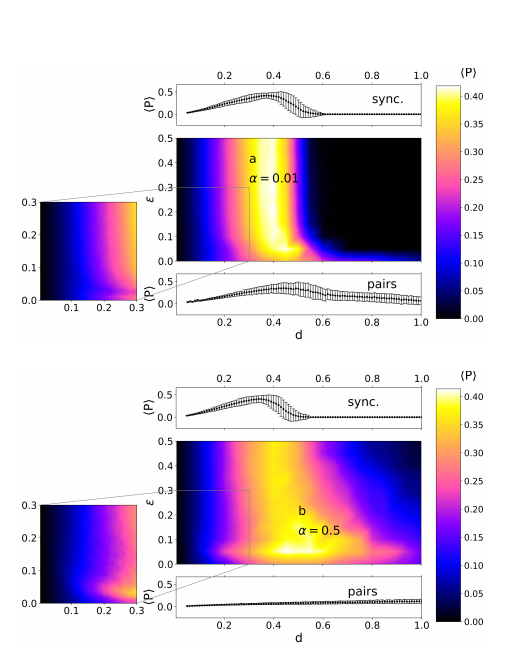}
\caption{Total average polarization of the system as a function of the basal confidence bound $d$ and the synchronization factor $\epsilon$ for (a) step length $\alpha = 0.01$, (b) $\alpha = 0.5$. Results obtained for a system of $N = 200$ agents and 100 simulations with different initial conditions. The inset plot shows a smaller range with more resolution. The upper and lower subpanels in each subfigure correspond to the synchronized and pairwise cases for the corresponding set of parameters.}
\label{Fig:polarization}       
\end{figure}

When analyzing the lower range of $\epsilon$ in detail (refer to Fig. \ref{Fig:polarization} - inset), we find a non-monotonic behavior. For instance, following the vertical line at $d = 0.2$, we see that the average polarization is $\langle P \rangle \approx 200$ for $\epsilon = 0$, decreases to a minimum of $\langle P \rangle \approx 75$ around $\epsilon = 0.02$, and rises again to approximately $\langle P \rangle \approx 200$ at $\epsilon = 0.1$. Beyond $\epsilon = 0.1$, the average final polarization primarily depends on $d$ only.

For $\epsilon > 0.05$, the system achieves consensus if $d > 0.6$, resulting in zero average polarization. When $\epsilon$ is below this threshold, the average polarization value $\langle P \rangle$ decreases with both $d$ and $\epsilon$.

Note that low polarization values at both low $d$ correspond to a highly fragmented outcome, while at high $d$, they indicate a consensus.
%
%
\subsubsection{High $\alpha$}\label{Susubsbec:polarization_high_alpha}

For large step length, polarization presents a different behavior, especially noticeable for low $\epsilon$ and high $d$, where it exhibits greater values compared to scenarios with smaller $\alpha$. For instance, taking the vertical line $d = 0.6$ (Fig. \ref{Fig:polarization} (b)), we observe a peak at $\epsilon \approx 0.05$. Additionally, the previously observed non-monotonic behavior for small $\epsilon$ and $0.1 < d < 0.3$ is inverted. In the vertical line at $d = 0.3$, we have a peak in the average polarization instead of a valley at a slightly larger $\epsilon \approx 0.05$. The peak across all $\epsilon > 0.1$ along $d = 0.4$ persists.

The results in the limit of high $\epsilon$ no longer converge to the synchronized case. With a larger step length, the initial steps, where the system has not synchronized yet, become crucial, inducing fragmentation, and thereby increasing polarization. This effect is prevented only when the system is fully synchronized from the beginning, leading to consensus.

This transition, more abrupt for the synchronized case, as usual, is absent in the pairwise scenario."
%
%
\subsubsection{Polarization peak}\label{Subsubsec:polarization_peak}

\begin{figure}[h!]
\centering
\includegraphics[width=1.0\linewidth]{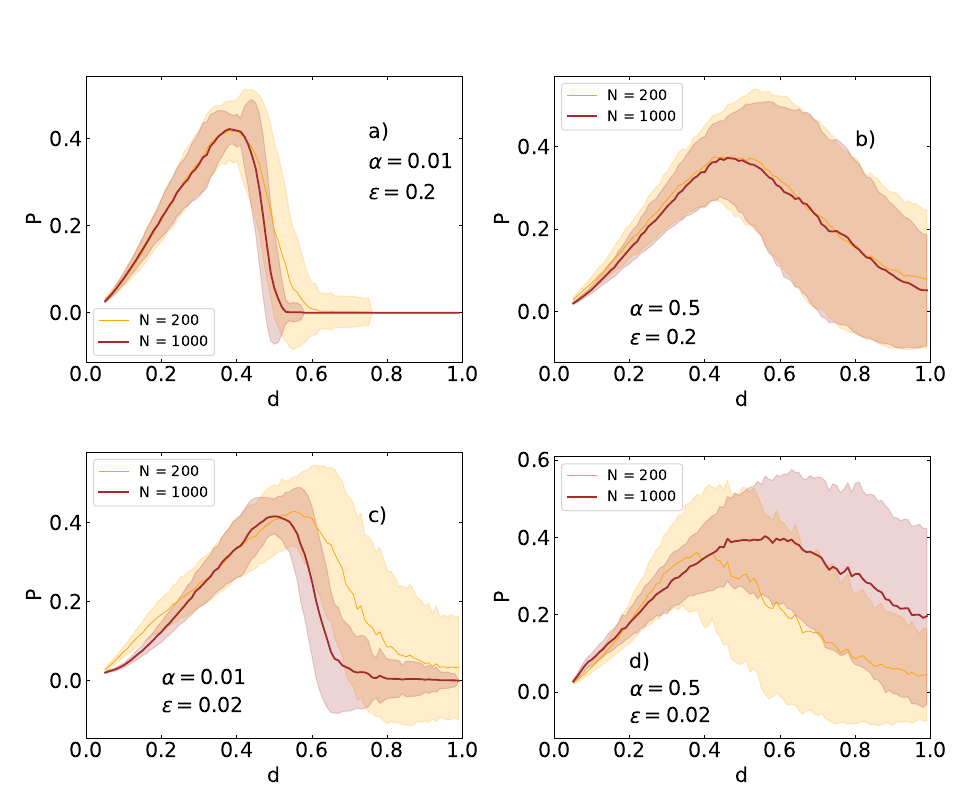}
\caption{Total polarization of the system as a function of the basal confidence bound $d$ for (a) step $\alpha = 0.01$ and coupling factor $\epsilon = 0.2$, (b) $\alpha = 0.5$ and coupling factor $\epsilon = 0.2$, (c) step $\alpha = 0.01$ and coupling factor $\epsilon = 0.02$, and (d) $\alpha = 0.5$ and coupling factor $\epsilon = 0.02$. Results obtained for a system of $N = 200$ agents (thin light line) and $N = 1000$ agents (thick dark line) and 300 simulations with different initial conditions.}
\label{Fig:polarization_total}       
\end{figure}

As observed, the system's global polarization peak strongly correlates with the bipartisan peak. This correlation becomes clearer when we compare Figures \ref{Fig:swap_d} and \ref{Fig:polarization_total}. With low $\alpha$, an increase in agent density sharpens this transition, as is typical in second-order transitions. When $\alpha$ increases, heightened fluctuations and disruptions in clusters lead to a scenario where $N$ no longer influences the average total polarization at low $\epsilon$. Moreover, a larger step smoothens the transition at high $\epsilon$.

%
%
%
\subsubsection{Polarization per outcome $\alpha$}\label{Subsubsec:polarization_clusters}
\begin{figure}[h!]
\centering
\includegraphics[width=1.0\linewidth]{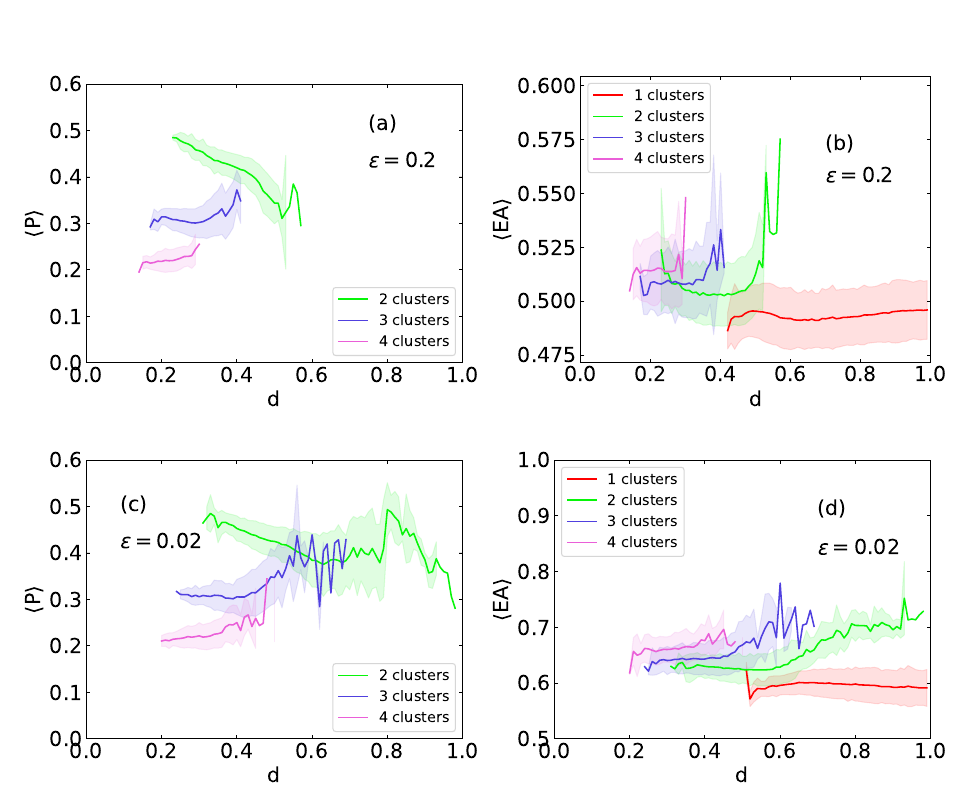}
\caption{Average polarization (left) and average final EA (right) of each outcome system as a function of the basal confidence bound $d$ Results are shown up to 4 clusters and are obtained for a system of $N = 1000$ agents, synchronization factor $\epsilon = 0.2$ (up) and $\epsilon = 0.02$ (down), step longitude $\alpha = 0.01$, and 300 simulations with different initial conditions.}
\label{Fig:polarization_clusters}       
\end{figure}

When examining average polarization levels based on outcomes (Fig. \ref{Fig:polarization_clusters}), the average polarization $\langle P \rangle$ increases proportionally with the basal confidence bound $d$ in scenarios where the system fragments into 3 or 4 clusters, in the same way the system increases the average level $\langle EA \rangle$. On the other hand, for the bipartisan outcome $\langle EA \rangle$ maintains the increasing tendency, while  $\langle P \rangle$ decreases.

Within the range of $d$ values where systems exhibit the coexistence of 2 and 3-cluster outcomes, realizations where the system significantly raises the $\langle EA \rangle$ result in more significant loss of connections among agents, causing the system to break into 3 clusters and resulting in decreased average polarization. In contrast, systems maintaining lower EA levels prevent fragmentation, leading to a bipartisan outcome and higher polarization levels. The final outcome is solely determined by initial conditions (position in the plane and first emitter).

At higher $\epsilon$ these tendencies become sharper (see Fig. \ref{Fig:polarization_clusters} (c) and (d)). Longer steps are not shown in Fig. \ref{Fig:polarization_clusters}) because the increased number of fluctuations outgrow the aforementioned behaviors.

%
\subsection{Bipartisanship}\label{Subsec:bipart}
In this subsection, we examine the bipartisan outcome, a common scenario in various social contexts such as presidential elections (e.g., U.S., France) and leadership contests within political parties. In Fig. \ref{Fig:polarization_clusters}, we analyze the polarization levels of realizations ending in 2 clusters depending on the model parameters. Polarization hinges on the relative sizes and distances between clusters. Since the two clusters tend to be similar in size (with fluctuations depending on parameter ranges), we are focusing on the distance between them.

\begin{figure}[h!]
\centering
\includegraphics[width=1.0\linewidth]{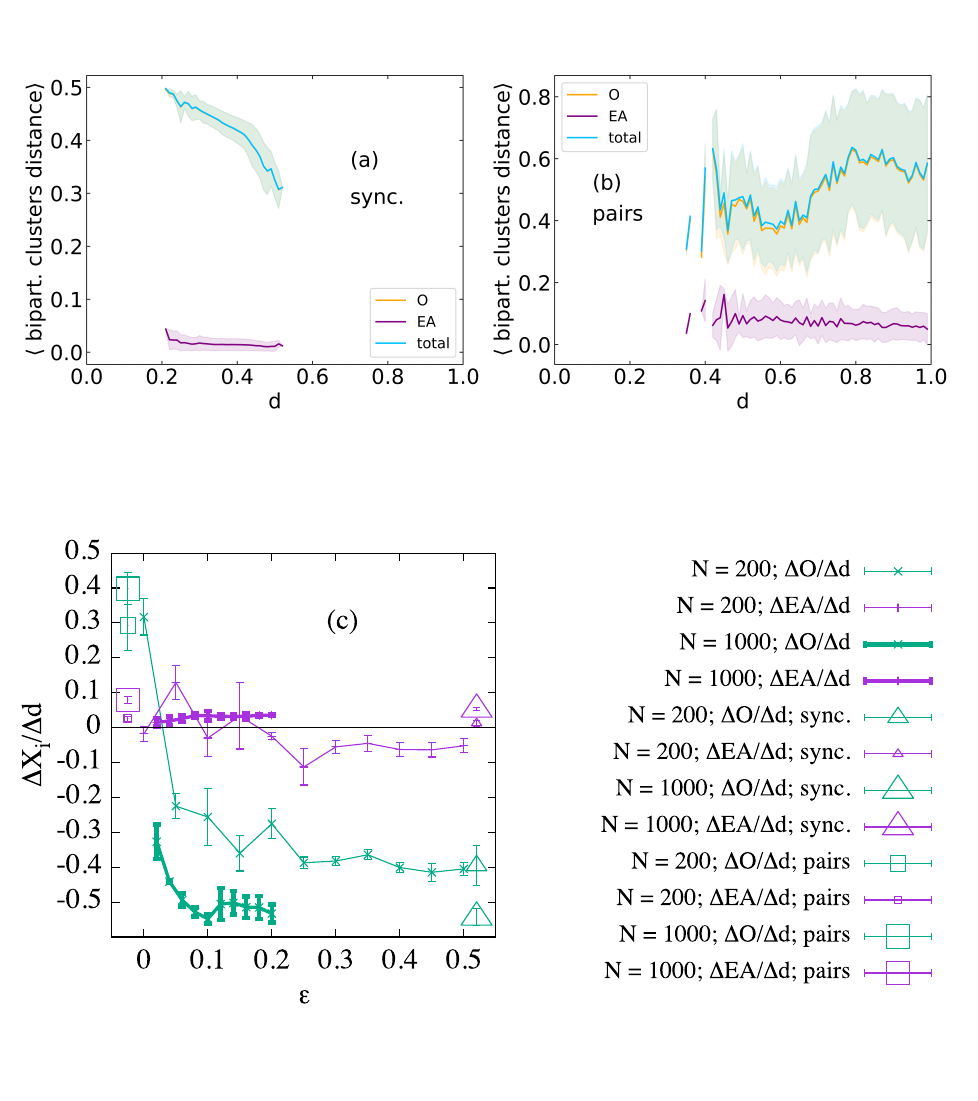}
\caption{Distance between the two clusters observed in realizations with a bipartisan outcome, shown on both the opinion axis O (in light blue) and the emotional arousal axis EA (in purple), plotted against the basal confidence bound $d$ for both (a) the synchronized case and (b) the pairwise case. (c) Slopes derived from linear regressions fitted to the O distance (in light blue) and the EA distance (in purple) between these clusters against $d$, accompanied by their respective errors, as a function of $\epsilon$. The results are obtained for step length $\alpha = 0.01$, and systems with $N = 1000$ agents for (a) and (b), and $N = 200, 1000$ agents in (c), conducting 300 repetitions with different initial conditions for each set of parameters. Values of $d$ with less than 10 percent of repetitions ending in bipartisanship were discarded for the linear regression.}
\label{Fig:bipart}       
\end{figure}

We present results for a step length $\alpha = 0.01$ because larger $\alpha$ blurs effects due to increased fluctuations. 
For larger steps, the initial messages become crucial, especially when emitters possess high initial $EA$, since they may draw some agents upward in the O-EA plane but not others. When $d$ is high this fosters bipartisanship over consensus, and clusters generally exhibit different EAs, contingent on the first emitters on each side of the opinion space, who attract agents to each cluster, and their respective $EA$ levels. However, for small $\alpha$, the vertical distance on the emotional arousal axis fluctuates around zero, with fluctuations decaying approximately as $1/\sqrt{N}$ (Fig. \ref{Fig:bipart} (a) and (b)).  

On the other hand, the distance on the opinion axis for small $\alpha$ depends on the synchronization level. In the fully synchronized case (Fig. \ref{Fig:bipart} (a)), the distance decreases with the basal confidence bound $d$ as larger confidence bounds bring clusters closer. In contrast, with pairwise interactions (Fig. \ref{Fig:bipart} (b)), it increases with $d$, because of the increased $EA$ levels reached in this case.

To explore slope changes, we fitted linear regressions to the data for both the synchronized and the pairwise case, and across various $\epsilon$ values (Fig. \ref{Fig:bipart} (c)). Similar to the results we expose in previous sections, sequential case results with $\epsilon = 0$ tend to converge to pairwise results, (more accurately for $N = 200$), while results for larger $\epsilon$ converge to synchronized case results. 

The emotional arousal axis distance maintains a flat slope in all cases, with fewer fluctuations for larger $N$. In contrast, the opinion axis distance undergoes a transition from positive to negative slopes around an $\epsilon$ transition value of approximately 0.05. The slope stabilizes for $\epsilon > 0.2$, matching synchronized case values. Slopes have larger absolute values and fewer fluctuations for larger $N$. At the transition value, local synchronization is achieved fast enough to have entire clusters firing at once early in the dynamics, when the system still forms a single connected component, which contributes to a cohesive approach in both the opinion and the EA axis, leading to lower levels of EA and closer clusters in the bipartisan outcome, so even in bipartisanship.

\section{Conclusions} \label{conclusions}
%
%
Our study reveals different stationary states for the proposed system, depending on the parameters set and the initial conditions. In general, we observe that the dynamics leads to an increment in the average of emotional arousal level of the system and a fragmentation of the initial number of clusters into several final opinion groups. However, for some specific set of parameters, a less fragmented situation is more probable. The synchronization factor $\epsilon$ significantly influences these aspects, reinforcing their effects when it is small and mitigating their intensity when is large.

Total synchronization emerges as a unifying factor that not only facilitates consensus and prevents fragmentation at lower basal confidence bound values, but also eliminates fluctuations, the outcome becoming more predictable, and sharpens transitions between different outcomes. Fully synchronized systems even reach consensus at lower confidence bounds for larger step longitudes. However, this scenario poses challenges in replicating real-world debates effectively, while it can be of interest in the design of deliberative processes. Synchronization, above all factors, appears as the most favorable to consensus attainment. Furthermore, the position of the consensus cluster has a lower EA than the initial average $\langle EA \rangle$. Additionally, in the synchronized case there is a narrow range of the basal confidence bound $d$, around $0.4$, within which the outcome is always bipartisanship regardless of initial conditions, the exact position of this $d$ range shifts slightly with $N$ and $\alpha$ towards lower $d$. Furthermore, synchronization not only fosters consensus but also diminishes opinion distance between clusters as $d$ increases in bipartisan outcomes. In contrast, in pairwise and sequential cases, this distance expands with $d$. 

Phase diagrams for $d$ and $\epsilon$ show more complicated profile demonstrating a more intrincate interplay between these two parameters than between $\alpha$ and $\epsilon$. A larger $\alpha$ disrupts transitions toward consensus, except for synchronized systems where in contrast larger $\alpha$ achieve consensus at lower $d$.

Agent density plays a relevant role in shaping the distribution of the system's basins of attraction as well. Having people holding opinions in between other people builds more links in the system and as a consequence higher agent densities tend to diminish the final number of opinion groups, particularly within parameter regions that lead to fragmentation into multiple clusters (low $\epsilon$, low $d$, and high $\alpha$)

The system is very dependent on the initial positions of the agents in the O-EA plane and on the relative position of the first emitter. 

The emotional arousal dimension enhances the relevance of small opinion groups, called wings. For instance, the presence of highly emotional minorities, although filtered, notably impacts final levels of emotional arousal, because they are capable of triggering a general increase in $\langle EA \rangle$ before segregating from the bigger groups. Small opinion groups, which used to appear only in the extremes of the opinion segment for the unidimensional model can now emerge in moderate opinions, especially for large step lengths. They also appear with lower levels of emotional arousal for small $\alpha$.

High $\alpha$ and low $d$ scenarios consistently result in highly fragmented situations, but in general fragmentation is intricately influenced by interaction patterns related to synchronization and step length per interaction. In practical scenarios, these magnitudes would reflect intrinsic attributes, different for each agent. However, as an initial approximation, considering the mean values of participating agents in the modeled scenario suffices. Navigating the interplay between various system parameters becomes complicated, particularly in promoting consensus within deliberative spaces. Achieving consensus can be a sensitive endeavor due to multiple influencing factors.

Our model, designed to simulate deliberative spaces, involves agents capable of potential interactions without any specific underlying network topology affecting their interaction patterns, which are conditioned only by the confidence interval. The model could be applied to describe opinion formation in debating chambers, representing hours-long deliberations with presumably short displacements after a round of interactions. Another potential scenario involves extended online discussions on online platforms, occurring over months, thus allowing again to consider a complete underlying graph. Larger alpha values probably model more effectively the latter situation. 

Another viewpoint on interpreting the parameters of the model pertains to the level of fragmentation or the number of active parties during country elections. A greater number of parties often correlates with longer step lengths contributing to opinion fragmentation, except in fully synchronized cases. Similarly, reducing synchronization or a decline in the basal confidence bound may increase the final number of opinion groups. The actual scenario likely results from a blend of multiple contributing factors.

%

\vspace*{0.5cm}
\begin{acknowledgments}
The authors acknowledge support from Spanish Grant No. PID2021-128005NB-C22, funded by
MCIN/AEI/10.13039/501100011033 and “ERDF A way of making Europe,” and from Generalitat de Catalunya (2021SGR00856).
I.F.’s work has been supported by Grant No. PRE2019-090279 (MCIN/AEI/10.13039/501100011033).
\end{acknowledgments}


\bibliographystyle{apsrev4-1}

\bibliography{ifo_bib}

\begin{thebibliography}{26}%
\makeatletter
\providecommand \@ifxundefined [1]{%
 \@ifx{#1\undefined}
}%
\providecommand \@ifnum [1]{%
 \ifnum #1\expandafter \@firstoftwo
 \else \expandafter \@secondoftwo
 \fi
}%
\providecommand \@ifx [1]{%
 \ifx #1\expandafter \@firstoftwo
 \else \expandafter \@secondoftwo
 \fi
}%
\providecommand \natexlab [1]{#1}%
\providecommand \enquote  [1]{``#1''}%
\providecommand \bibnamefont  [1]{#1}%
\providecommand \bibfnamefont [1]{#1}%
\providecommand \citenamefont [1]{#1}%
\providecommand \href@noop [0]{\@secondoftwo}%
\providecommand \href [0]{\begingroup \@sanitize@url \@href}%
\providecommand \@href[1]{\@@startlink{#1}\@@href}%
\providecommand \@@href[1]{\endgroup#1\@@endlink}%
\providecommand \@sanitize@url [0]{\catcode `\\12\catcode `\$12\catcode
  `\&12\catcode `\#12\catcode `\^12\catcode `\_12\catcode `\%12\relax}%
\providecommand \@@startlink[1]{}%
\providecommand \@@endlink[0]{}%
\providecommand \url  [0]{\begingroup\@sanitize@url \@url }%
\providecommand \@url [1]{\endgroup\@href {#1}{\urlprefix }}%
\providecommand \urlprefix  [0]{URL }%
\providecommand \Eprint [0]{\href }%
\providecommand \doibase [0]{http://dx.doi.org/}%
\providecommand \selectlanguage [0]{\@gobble}%
\providecommand \bibinfo  [0]{\@secondoftwo}%
\providecommand \bibfield  [0]{\@secondoftwo}%
\providecommand \translation [1]{[#1]}%
\providecommand \BibitemOpen [0]{}%
\providecommand \bibitemStop [0]{}%
\providecommand \bibitemNoStop [0]{.\EOS\space}%
\providecommand \EOS [0]{\spacefactor3000\relax}%
\providecommand \BibitemShut  [1]{\csname bibitem#1\endcsname}%
\let\auto@bib@innerbib\@empty
\bibitem [{\citenamefont {Liggett}(2013)}]{liggett2013stochastic}%
  \BibitemOpen
  \bibfield  {author} {\bibinfo {author} {\bibfnamefont {T.~M.}\ \bibnamefont
  {Liggett}},\ }\href@noop {} {\emph {\bibinfo {title} {Stochastic interacting
  systems: contact, voter and exclusion processes}}},\ Vol.\ \bibinfo {volume}
  {324}\ (\bibinfo  {publisher} {springer science \& Business Media},\ \bibinfo
  {year} {2013})\BibitemShut {NoStop}%
\bibitem [{\citenamefont {Lorenz}(2007)}]{lorenz2007continuous}%
  \BibitemOpen
  \bibfield  {author} {\bibinfo {author} {\bibfnamefont {J.}~\bibnamefont
  {Lorenz}},\ }\href@noop {} {\bibfield  {journal} {\bibinfo  {journal}
  {International Journal of Modern Physics C}\ }\textbf {\bibinfo {volume}
  {18}},\ \bibinfo {pages} {1819} (\bibinfo {year} {2007})}\BibitemShut
  {NoStop}%
\bibitem [{\citenamefont {Castellano}\ \emph {et~al.}(2009)\citenamefont
  {Castellano}, \citenamefont {Fortunato},\ and\ \citenamefont
  {Loreto}}]{Castellano_2009}%
  \BibitemOpen
  \bibfield  {author} {\bibinfo {author} {\bibfnamefont {C.}~\bibnamefont
  {Castellano}}, \bibinfo {author} {\bibfnamefont {S.}~\bibnamefont
  {Fortunato}}, \ and\ \bibinfo {author} {\bibfnamefont {V.}~\bibnamefont
  {Loreto}},\ }\href {\doibase 10.1103/revmodphys.81.591} {\bibfield  {journal}
  {\bibinfo  {journal} {Rev. Mod. Phys.}\ }\textbf {\bibinfo {volume} {81}},\
  \bibinfo {pages} {591–646} (\bibinfo {year} {2009})}\BibitemShut {NoStop}%
\bibitem [{\citenamefont {Hegselmann}\ and\ \citenamefont
  {Krause}(2002)}]{Hegselmann_2002}%
  \BibitemOpen
  \bibfield  {author} {\bibinfo {author} {\bibfnamefont {R.}~\bibnamefont
  {Hegselmann}}\ and\ \bibinfo {author} {\bibfnamefont {U.}~\bibnamefont
  {Krause}},\ }\href {http://jasss.soc.surrey.ac.uk/5/3/2.html} {\bibfield
  {journal} {\bibinfo  {journal} {J. Artif. Soc. Soc. Simul.}\ }\textbf
  {\bibinfo {volume} {5}},\ \bibinfo {pages} {2} (\bibinfo {year}
  {2002})}\BibitemShut {NoStop}%
\bibitem [{\citenamefont {Deffuant}\ \emph {et~al.}(2000)\citenamefont
  {Deffuant}, \citenamefont {Neau}, \citenamefont {Amblard},\ and\
  \citenamefont {Weisbuch}}]{Deffuant_2000}%
  \BibitemOpen
  \bibfield  {author} {\bibinfo {author} {\bibfnamefont {G.}~\bibnamefont
  {Deffuant}}, \bibinfo {author} {\bibfnamefont {D.}~\bibnamefont {Neau}},
  \bibinfo {author} {\bibfnamefont {F.}~\bibnamefont {Amblard}}, \ and\
  \bibinfo {author} {\bibfnamefont {G.}~\bibnamefont {Weisbuch}},\ }\href
  {https://EconPapers.repec.org/RePEc:wsi:acsxxx:v:03:y:2000:i:01n04:n:s0219525900000078}
  {\bibfield  {journal} {\bibinfo  {journal} {Adv. in Compl. Syst.}\ }\textbf
  {\bibinfo {volume} {3}},\ \bibinfo {pages} {87} (\bibinfo {year}
  {2000})}\BibitemShut {NoStop}%
\bibitem [{\citenamefont {Flache}\ \emph {et~al.}(2017)\citenamefont {Flache},
  \citenamefont {M{\"a}s}, \citenamefont {Feliciani}, \citenamefont
  {Chattoe-Brown}, \citenamefont {Deffuant}, \citenamefont {Huet},\ and\
  \citenamefont {Lorenz}}]{flache2017models}%
  \BibitemOpen
  \bibfield  {author} {\bibinfo {author} {\bibfnamefont {A.}~\bibnamefont
  {Flache}}, \bibinfo {author} {\bibfnamefont {M.}~\bibnamefont {M{\"a}s}},
  \bibinfo {author} {\bibfnamefont {T.}~\bibnamefont {Feliciani}}, \bibinfo
  {author} {\bibfnamefont {E.}~\bibnamefont {Chattoe-Brown}}, \bibinfo {author}
  {\bibfnamefont {G.}~\bibnamefont {Deffuant}}, \bibinfo {author}
  {\bibfnamefont {S.}~\bibnamefont {Huet}}, \ and\ \bibinfo {author}
  {\bibfnamefont {J.}~\bibnamefont {Lorenz}},\ }\href@noop {} {\bibfield
  {journal} {\bibinfo  {journal} {Jasss-The journal of artificial societies and
  social simulation}\ }\textbf {\bibinfo {volume} {20}},\ \bibinfo {pages} {2}
  (\bibinfo {year} {2017})}\BibitemShut {NoStop}%
\bibitem [{\citenamefont {Carpentras}(2023)}]{carpentras2023we}%
  \BibitemOpen
  \bibfield  {author} {\bibinfo {author} {\bibfnamefont {D.}~\bibnamefont
  {Carpentras}},\ }\href@noop {} {\bibfield  {journal} {\bibinfo  {journal}
  {Review of Artificial Societies and Social Simulations}\ } (\bibinfo {year}
  {2023})}\BibitemShut {NoStop}%
\bibitem [{\citenamefont {Sobkowicz}(2009)}]{sobkowicz2009modelling}%
  \BibitemOpen
  \bibfield  {author} {\bibinfo {author} {\bibfnamefont {P.}~\bibnamefont
  {Sobkowicz}},\ }\href@noop {} {\bibfield  {journal} {\bibinfo  {journal}
  {Journal of Artificial Societies and Social Simulation}\ }\textbf {\bibinfo
  {volume} {12}},\ \bibinfo {pages} {11} (\bibinfo {year} {2009})}\BibitemShut
  {NoStop}%
\bibitem [{\citenamefont {Helbing}\ \emph {et~al.}(2023)\citenamefont
  {Helbing}, \citenamefont {Mahajan}, \citenamefont {Fricker}, \citenamefont
  {Musso}, \citenamefont {Hausladen}, \citenamefont {Carissimo}, \citenamefont
  {Carpentras}, \citenamefont {Stockinger}, \citenamefont {Sanchez-Vaquerizo},
  \citenamefont {Yang} \emph {et~al.}}]{helbing2023democracy}%
  \BibitemOpen
  \bibfield  {author} {\bibinfo {author} {\bibfnamefont {D.}~\bibnamefont
  {Helbing}}, \bibinfo {author} {\bibfnamefont {S.}~\bibnamefont {Mahajan}},
  \bibinfo {author} {\bibfnamefont {R.~H.}\ \bibnamefont {Fricker}}, \bibinfo
  {author} {\bibfnamefont {A.}~\bibnamefont {Musso}}, \bibinfo {author}
  {\bibfnamefont {C.~I.}\ \bibnamefont {Hausladen}}, \bibinfo {author}
  {\bibfnamefont {C.}~\bibnamefont {Carissimo}}, \bibinfo {author}
  {\bibfnamefont {D.}~\bibnamefont {Carpentras}}, \bibinfo {author}
  {\bibfnamefont {E.}~\bibnamefont {Stockinger}}, \bibinfo {author}
  {\bibfnamefont {J.~A.}\ \bibnamefont {Sanchez-Vaquerizo}}, \bibinfo {author}
  {\bibfnamefont {J.~C.}\ \bibnamefont {Yang}},  \emph {et~al.},\ }\href@noop
  {} {\bibfield  {journal} {\bibinfo  {journal} {Journal of Computational
  Science}\ }\textbf {\bibinfo {volume} {71}},\ \bibinfo {pages} {102061}
  (\bibinfo {year} {2023})}\BibitemShut {NoStop}%
\bibitem [{\citenamefont {Barandiaran}\ \emph {et~al.}(2024)\citenamefont
  {Barandiaran}, \citenamefont {Calleja-López}, \citenamefont {Monterde},\
  and\ \citenamefont {Romero}}]{barandiaran2024decidim}%
  \BibitemOpen
  \bibfield  {author} {\bibinfo {author} {\bibfnamefont {X.~E.}\ \bibnamefont
  {Barandiaran}}, \bibinfo {author} {\bibfnamefont {A.}~\bibnamefont
  {Calleja-López}}, \bibinfo {author} {\bibfnamefont {A.}~\bibnamefont
  {Monterde}}, \ and\ \bibinfo {author} {\bibfnamefont {C.}~\bibnamefont
  {Romero}},\ }\href {\doibase 10.1007/978-3-031-50784-7} {\emph {\bibinfo
  {title} {Decidim, a Technopolitical Network for Participatory Democracy:
  Philosophy, Practice and Autonomy of a Collective Platform in the Age of
  Digital Intelligence}}}\ (\bibinfo  {publisher} {Springer Nature
  Switzerland},\ \bibinfo {year} {2024})\BibitemShut {NoStop}%
\bibitem [{\citenamefont {Urbig}\ \emph {et~al.}(2008)\citenamefont {Urbig},
  \citenamefont {Lorenz},\ and\ \citenamefont {Herzberg}}]{urbig2008opinion}%
  \BibitemOpen
  \bibfield  {author} {\bibinfo {author} {\bibfnamefont {D.}~\bibnamefont
  {Urbig}}, \bibinfo {author} {\bibfnamefont {J.}~\bibnamefont {Lorenz}}, \
  and\ \bibinfo {author} {\bibfnamefont {H.}~\bibnamefont {Herzberg}},\
  }\href@noop {} {\bibfield  {journal} {\bibinfo  {journal} {Journal of
  Artificial Societies and Social Simulation}\ }\textbf {\bibinfo {volume}
  {11}},\ \bibinfo {pages} {4} (\bibinfo {year} {2008})}\BibitemShut {NoStop}%
\bibitem [{\citenamefont
  {Veefkind~Castell{\'o}}(2022)}]{veefkind2022consensus}%
  \BibitemOpen
  \bibfield  {author} {\bibinfo {author} {\bibfnamefont {J.}~\bibnamefont
  {Veefkind~Castell{\'o}}},\ }\href
  {https://diposit.ub.edu/dspace/bitstream/2445/201886/1/VEEFKIND\%20CASTELL\%C3\%93\%20JORIS\_7934134.pdf}
  {\enquote {\bibinfo {title} {Consensus formation in the deffuant model of
  opinion dynamics},}\ } (\bibinfo {year} {2022}),\ \bibinfo {note} {bachelor's
  thesis}\BibitemShut {NoStop}%
\bibitem [{\citenamefont {Piedrahita}\ \emph {et~al.}(2013)\citenamefont
  {Piedrahita}, \citenamefont {Borge-Holthoefer}, \citenamefont {Moreno},\ and\
  \citenamefont {Arenas}}]{piedrahita2013modeling}%
  \BibitemOpen
  \bibfield  {author} {\bibinfo {author} {\bibfnamefont {P.}~\bibnamefont
  {Piedrahita}}, \bibinfo {author} {\bibfnamefont {J.}~\bibnamefont
  {Borge-Holthoefer}}, \bibinfo {author} {\bibfnamefont {Y.}~\bibnamefont
  {Moreno}}, \ and\ \bibinfo {author} {\bibfnamefont {A.}~\bibnamefont
  {Arenas}},\ }\href@noop {} {\bibfield  {journal} {\bibinfo  {journal}
  {Europhysics letters}\ }\textbf {\bibinfo {volume} {104}},\ \bibinfo {pages}
  {48004} (\bibinfo {year} {2013})}\BibitemShut {NoStop}%
\bibitem [{\citenamefont {Schweitzer}\ \emph {et~al.}(2020)\citenamefont
  {Schweitzer}, \citenamefont {Krivachy},\ and\ \citenamefont
  {Garcia}}]{Schweitzer_2020}%
  \BibitemOpen
  \bibfield  {author} {\bibinfo {author} {\bibfnamefont {F.}~\bibnamefont
  {Schweitzer}}, \bibinfo {author} {\bibfnamefont {T.}~\bibnamefont
  {Krivachy}}, \ and\ \bibinfo {author} {\bibfnamefont {D.}~\bibnamefont
  {Garcia}},\ }\href {\doibase 10.1155/2020/5282035} {\bibfield  {journal}
  {\bibinfo  {journal} {Complexity}\ }\textbf {\bibinfo {volume} {2020}},\
  \bibinfo {pages} {1} (\bibinfo {year} {2020})}\BibitemShut {NoStop}%
\bibitem [{\citenamefont {Sobkowicz}(2015)}]{Sobkowicz_2015}%
  \BibitemOpen
  \bibfield  {author} {\bibinfo {author} {\bibfnamefont {P.}~\bibnamefont
  {Sobkowicz}},\ }\href {\doibase 10.3389/fphy.2015.00017} {\bibfield
  {journal} {\bibinfo  {journal} {Front. Phys.}\ }\textbf {\bibinfo {volume}
  {3}},\ \bibinfo {pages} {17} (\bibinfo {year} {2015})}\BibitemShut {NoStop}%
\bibitem [{\citenamefont {Sobkowicz}(2012)}]{sobkowicz2012discrete}%
  \BibitemOpen
  \bibfield  {author} {\bibinfo {author} {\bibfnamefont {P.}~\bibnamefont
  {Sobkowicz}},\ }\href@noop {} {\bibfield  {journal} {\bibinfo  {journal}
  {PLoS One}\ } (\bibinfo {year} {2012})}\BibitemShut {NoStop}%
\bibitem [{\citenamefont {Colaiori}\ \emph {et~al.}(2016)\citenamefont
  {Colaiori}, \citenamefont {Castellano}, \citenamefont {Caccioli},\ and\
  \citenamefont {Vivo}}]{colaiori2016consensus}%
  \BibitemOpen
  \bibfield  {author} {\bibinfo {author} {\bibfnamefont {F.}~\bibnamefont
  {Colaiori}}, \bibinfo {author} {\bibfnamefont {C.}~\bibnamefont
  {Castellano}}, \bibinfo {author} {\bibfnamefont {F.}~\bibnamefont
  {Caccioli}}, \ and\ \bibinfo {author} {\bibfnamefont {P.}~\bibnamefont
  {Vivo}},\ }\href {\doibase 10.1088/1742-5468/2016/03/033401} {\bibfield
  {journal} {\bibinfo  {journal} {J. Stat. Mech. Theory Exp.}\ }\textbf
  {\bibinfo {volume} {2016}},\ \bibinfo {pages} {033401} (\bibinfo {year}
  {2016})}\BibitemShut {NoStop}%
\bibitem [{\citenamefont {Prignano}\ \emph {et~al.}(2013)\citenamefont
  {Prignano}, \citenamefont {Sagarra},\ and\ \citenamefont
  {D{\'\i}az-Guilera}}]{prignano2013tuning}%
  \BibitemOpen
  \bibfield  {author} {\bibinfo {author} {\bibfnamefont {L.}~\bibnamefont
  {Prignano}}, \bibinfo {author} {\bibfnamefont {O.}~\bibnamefont {Sagarra}}, \
  and\ \bibinfo {author} {\bibfnamefont {A.}~\bibnamefont
  {D{\'\i}az-Guilera}},\ }\href@noop {} {\bibfield  {journal} {\bibinfo
  {journal} {Physical review letters}\ }\textbf {\bibinfo {volume} {110}},\
  \bibinfo {pages} {114101} (\bibinfo {year} {2013})}\BibitemShut {NoStop}%
\bibitem [{\citenamefont {Bottani}(1996)}]{bottani1996synchronization}%
  \BibitemOpen
  \bibfield  {author} {\bibinfo {author} {\bibfnamefont {S.}~\bibnamefont
  {Bottani}},\ }\href {\doibase 10.1103/PhysRevE.54.2334} {\bibfield  {journal}
  {\bibinfo  {journal} {Phys. Rev. E}\ }\textbf {\bibinfo {volume} {54}},\
  \bibinfo {pages} {2334} (\bibinfo {year} {1996})}\BibitemShut {NoStop}%
\bibitem [{\citenamefont {Mobilia}\ \emph {et~al.}(2007)\citenamefont
  {Mobilia}, \citenamefont {Petersen},\ and\ \citenamefont
  {Redner}}]{mobilia2007role}%
  \BibitemOpen
  \bibfield  {author} {\bibinfo {author} {\bibfnamefont {M.}~\bibnamefont
  {Mobilia}}, \bibinfo {author} {\bibfnamefont {A.}~\bibnamefont {Petersen}}, \
  and\ \bibinfo {author} {\bibfnamefont {S.}~\bibnamefont {Redner}},\
  }\href@noop {} {\bibfield  {journal} {\bibinfo  {journal} {Journal of
  Statistical Mechanics: Theory and Experiment}\ }\textbf {\bibinfo {volume}
  {2007}},\ \bibinfo {pages} {P08029} (\bibinfo {year} {2007})}\BibitemShut
  {NoStop}%
\bibitem [{Note1()}]{Note1}%
  \BibitemOpen
  \bibinfo {note} {While it is preferable to use therm \protect \textit
  {opinion group} to refer to a group of people that share a point of view, and
  the therm \protect \textit {cluster} to refer to agents with the same value
  of the variable opinion in the model, in this work we use the two therms
  interchangeably}\BibitemShut {NoStop}%
\bibitem [{\citenamefont {G{\'o}mez-Serrano}\ \emph {et~al.}(2012)\citenamefont
  {G{\'o}mez-Serrano}, \citenamefont {Graham},\ and\ \citenamefont
  {Le~Boudec}}]{gomez2012bounded}%
  \BibitemOpen
  \bibfield  {author} {\bibinfo {author} {\bibfnamefont {J.}~\bibnamefont
  {G{\'o}mez-Serrano}}, \bibinfo {author} {\bibfnamefont {C.}~\bibnamefont
  {Graham}}, \ and\ \bibinfo {author} {\bibfnamefont {J.-Y.}\ \bibnamefont
  {Le~Boudec}},\ }\href@noop {} {\bibfield  {journal} {\bibinfo  {journal}
  {Mathematical Models and Methods in Applied Sciences}\ }\textbf {\bibinfo
  {volume} {22}},\ \bibinfo {pages} {1150007} (\bibinfo {year}
  {2012})}\BibitemShut {NoStop}%
\bibitem [{\citenamefont {Barandiaran}\ \emph {et~al.}(2020)\citenamefont
  {Barandiaran}, \citenamefont {Calleja-L{\'o}pez},\ and\ \citenamefont
  {Cozzo}}]{barandiaran2020defining}%
  \BibitemOpen
  \bibfield  {author} {\bibinfo {author} {\bibfnamefont {X.~E.}\ \bibnamefont
  {Barandiaran}}, \bibinfo {author} {\bibfnamefont {A.}~\bibnamefont
  {Calleja-L{\'o}pez}}, \ and\ \bibinfo {author} {\bibfnamefont
  {E.}~\bibnamefont {Cozzo}},\ }\href@noop {} {\bibfield  {journal} {\bibinfo
  {journal} {Frontiers in Psychology}\ }\textbf {\bibinfo {volume} {11}},\
  \bibinfo {pages} {1549} (\bibinfo {year} {2020})}\BibitemShut {NoStop}%
\bibitem [{\citenamefont {Esteban}\ and\ \citenamefont
  {Ray}(1994)}]{Esteban_1994}%
  \BibitemOpen
  \bibfield  {author} {\bibinfo {author} {\bibfnamefont {J.~M.}\ \bibnamefont
  {Esteban}}\ and\ \bibinfo {author} {\bibfnamefont {D.}~\bibnamefont {Ray}},\
  }\href {\doibase 10.2307/2951734} {\bibfield  {journal} {\bibinfo  {journal}
  {Econometrica}\ }\textbf {\bibinfo {volume} {62}},\ \bibinfo {pages} {819}
  (\bibinfo {year} {1994})}\BibitemShut {NoStop}%
\bibitem [{\citenamefont {Kan}\ \emph {et~al.}(2023)\citenamefont {Kan},
  \citenamefont {Feng},\ and\ \citenamefont {Porter}}]{Kan_2022}%
  \BibitemOpen
  \bibfield  {author} {\bibinfo {author} {\bibfnamefont {U.}~\bibnamefont
  {Kan}}, \bibinfo {author} {\bibfnamefont {M.}~\bibnamefont {Feng}}, \ and\
  \bibinfo {author} {\bibfnamefont {M.~A.}\ \bibnamefont {Porter}},\ }\href
  {\doibase 10.1093/comnet/cnac055} {\bibfield  {journal} {\bibinfo  {journal}
  {Journal of Complex Networks}\ }\textbf {\bibinfo {volume} {11}},\ \bibinfo
  {pages} {415} (\bibinfo {year} {2023})}\BibitemShut {NoStop}%
\bibitem [{Note2()}]{Note2}%
  \BibitemOpen
  \bibinfo {note} {Https://github.com/Dialoguem/telegram-bot}\BibitemShut
  {NoStop}%
\end{thebibliography}%
\begin{appendix}
\onecolumngrid


\section{The algorithm}\label{App:algorithm}
\renewcommand{\theequation}{A-\arabic{equation}}
\setcounter{figure}{0}
\renewcommand{\thefigure}{A-\arabic{figure}}

%
\algnewcommand{\algorithmicand}{\textbf{ and }}
\algnewcommand{\algorithmicor}{\textbf{ or }}
\algnewcommand{\OR}{\algorithmicor}
\algnewcommand{\AND}{\algorithmicand}
\algnewcommand\TRUE{\textkeyword{\bf{True}}\space}
\algnewcommand{\var}{\texttt}

\begin{minipage}{\columnwidth}
\begin{algorithm}[H]
 \begin{algorithmic}
 \newcommand{\pseudotrue}{\textbf{True}}
 \label{mycode}
  \Require Number of agents $N$. A length step $\alpha$, a basal confidence bound $d$ (confidence bound at $EA$ = 0) and a synchronization factor $\epsilon$. A tolerance $\mu$ to consider stationarity.
  \Ensure Getting to one of the stationary attractors of the dynamics, given a sufficiently large $N_{steps}$
  \Statex
  \State set $step$ = 0
  \State set $step_{sync}$ = 0  
  
   \State set initial positions  ${\{\bf{O}}_i\}_{initial}$ and ${\{\bf{EA}}_i\}_{initial}$ uniformly at random in the range $[0, 1]$
   \State set initial phases ${\{\bf{\phi}}_i\}_{initial}$ uniformly at random in the range $[0, 1]$ (or ${\{\bf{\phi}}_i\}_{initial} = 1$ in the sync. case
  
\Repeat

   \State $\Delta \phi =  1-\phi_{max}$
   \State  Update all agents phases $ {\{ \bf{\phi}}_i \} =  {\{ \bf{\phi}}_i \}  + \Delta \phi $
      \Repeat
   \ForAll {$\phi_i$ $\phi_j$ with i, j $\in 1,...,N$}
   
     \If {${\bf{\phi}}_i = 1$}
     \State $dist_{ij} = \sqrt{(O_i - O_j)^2 + EA_i -EA_j)^2}$
     \If {$dist_{ij} < d$}
     \State $c = min\{ 1, \alpha/dist \} $
     \State ${\bf{O}_j} = {\bf{O}_j} + c{\bf{O}_j}$
      \State ${\bf{EA}_j} = {\bf{EA}_j} + c{\bf{EA}_j} $  
       \State ${\bf{\phi}}_j = \epsilon {\bf{\phi}}_j$ 
    \EndIf
     \EndIf
     \EndFor

    \Until {${\{\bf{\phi}}_i \ne 1$}
   \State $step = step + 1$
   \If{${\bf{\phi}}_i = {\bf{\phi}}_j$ $\forall$ i, j $\in$ same connected component} 
    \State $step_{sync} = step_{sync} + 1$
   \EndIf
  \Until {$\{\bf{\lvert \Delta O_i}\rvert \} , \, \{ \bf{\lvert \Delta EA_i} \rvert \} < \mu$}
  \State \Return { ${\{\bf{O}}_i\}_{final}$ and ${\{\bf{EA}}_i\}_{final}$, $step$ and $step_{sync}$ set of final positions states, number of steps when the simulation stops and number of step after local syncronization}
 \end{algorithmic}
 \caption{Performs an agent-based simulation using the broadcasting bounded confidence model with emotional arousal and synchronization dynamics.}
\end{algorithm}
\end{minipage}

%
%
\section{Wings}\label{App:wings}
\renewcommand{\theequation}{B-\arabic{equation}}
\setcounter{figure}{0}
\renewcommand{\thefigure}{B-\arabic{figure}}

\begin{figure}[h!]
\centering
\includegraphics[width=1.0\linewidth]{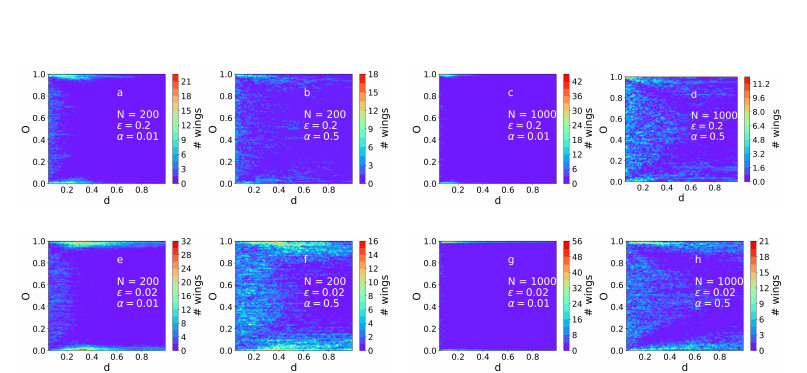}
\caption{Number of wings as a function of their opinion and the basal confidence bound $d$ for a coupling factor $\epsilon = 0.2$ (upper row) and $\epsilon = 0.02$ (lower row), system sizes $N = 200$ (a, b, e, f) and $N = 1000$ (c, d, g, h) and step longitude $\alpha = 0.01$ (a, c, e, g) and $\alpha = 0.5$ (b, d, f, h). Results for 300 simulations with different initial conditions.}
\label{Fig:wings_O}       
\end{figure}

In the one-dimensional model \cite{Deffuant_2000}, wings were disregarded due to their negligible role. However, in the model proposed in this study, wings assume greater significance, exhibiting a notable dependence on all parameters. Contrary to their mere appearance at the extremes of the opinion axis, these wings manifest not only at these endpoints but also emerge across various opinion positions, particularly prominent for low $d$ values, as shown in Fig. \ref{Fig:wings_O}. A larger step longitude accentuates the dispersion of wings along the opinion axis. For a system of $N = 200$ agents,  a small step longitude $\alpha$, and a high synchronization factor $\epsilon$ (Fig. \ref{Fig:wings_O} (a)), these wings tend to appear near the equilibrium points of clusters within the range $0.15 < d < 0.3$ (refer to Subsection \ref{Subsec:clust}). This suggests that these points are system attractors for the fully synchronized case, extending their significance beyond their role as preferred positions for the primary clusters.

\begin{figure}[h!]
\centering
\includegraphics[width=1.0\linewidth]{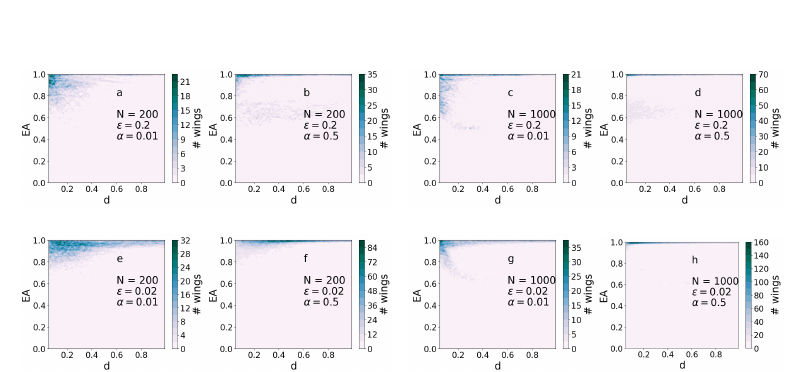}
\caption{Number of wings as a function of their emotional arousal and the basal confidence bound $d$ for a coupling factor $\epsilon = 0.2$ (upper row) and $\epsilon = 0.02$ (lower row), system sizes $N = 200$ (a, b, e, f) and $N = 1000$ (c, d, g, h) and step longitude $\alpha = 0.01$ (a, c, e, g) and $0.5$ (b, d, f, h). Results for 300 simulations with different initial conditions.}
\label{Fig:wings_EA}       
\end{figure}

In our model, wings have typically a large EA, and that's why they become isolated from the rest of the system. However, in some cases, wings can also have lower EA (see Fig. \ref{Fig:wings_EA}, especially in the parameter region with a clear predominance of a bipartisan outcome.

\section{The approximation for the synchronized case}\label{App:sync}
\renewcommand{\theequation}{C-\arabic{equation}}
\setcounter{figure}{0}
\renewcommand{\thefigure}{C-\arabic{figure}}

%
%
\begin{figure}[h!]
\centering
\includegraphics[width=1.0\linewidth]{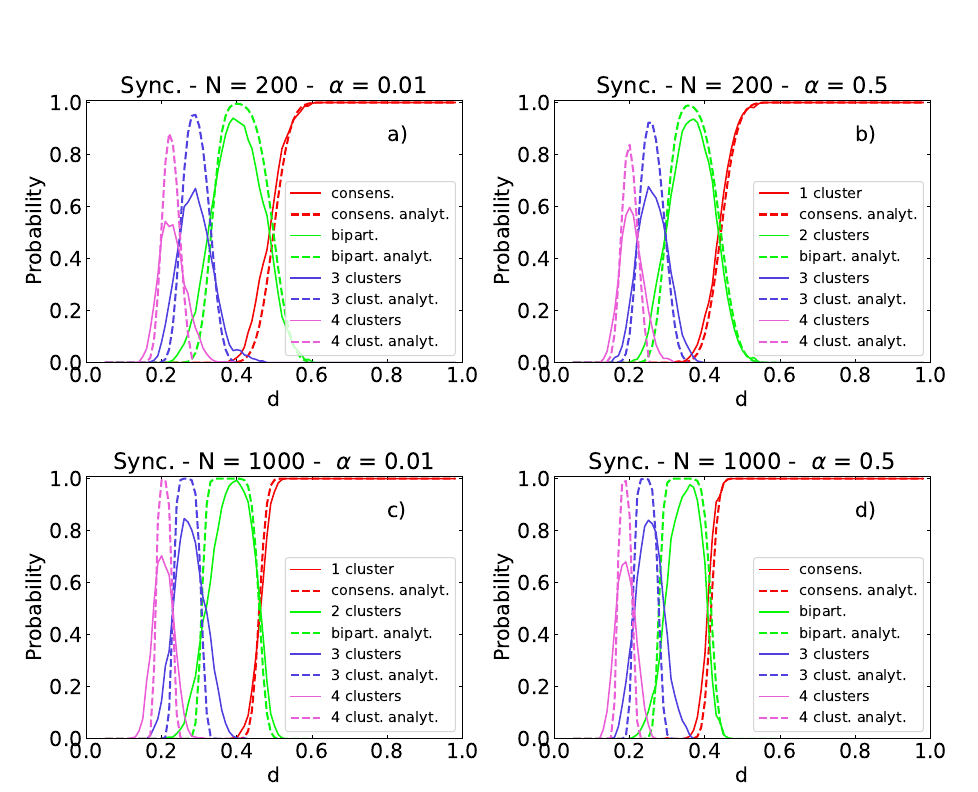}
\caption{Proportion of final scenarios as a function of the basal confidence bound $d$ for a) synchronized interactions, step longitude $\alpha = 0.01$ and system size $N = 200$, b) synchronized interactions, $\alpha = 0.5$ and $N = 200$, c) synchronized interactions, $\alpha = 0.01$ and $N = 1000$, d) step $\alpha = 0.5$ and  $N = 1000$. Results were obtained for 300 simulations with different initial conditions. Continuous lines correspond to simulations while dashed lines are for the semi-analytical approximation.}
\label{Fig:clust_sync_pairs_modified}     
\end{figure}
\begin{table*}
\caption{\label{Tab:sync} Table illustrating the minimum number of agents required to initially lie within each outcome's affectation area in order to estimate its probability, for different system sizes $N$ and step lengths $\alpha$.}
\begin{center}
\renewcommand\arraystretch{1.8}
\begin{tabular}{| >{\raggedright\arraybackslash}p{0.2\linewidth} | >{\raggedright\arraybackslash}p{0.2\linewidth} | >{\raggedright\arraybackslash}p{0.25\linewidth} |}
\hline
\textbf{N} & \textbf{$\alpha$} & \textbf{Sum lower bound Eq. \ref{Eq:sync}}\\  
\hline
200
&
0.01
&
$N/2$
\\\hline
200
&
0.5
&
$N/2 - N/18$
\\\hline
1000
&
0.01
&
$N/2 - N/25$
\\\hline
1000
&
0.5
&
$N/2 - N/12$
\\\hline
\end{tabular}
\end{center}
\end{table*}

We can improve the estimation for the different outcomes in the synchronized case if we modify the lower bound of the second sum in Eq. \ref{Eq:sync}, requiring a lower minimum number of agents to lie initially within the considered attraction area. Although we see that the analytical expression for this bound is an increasing function of $\alpha$ and $N$. In this Appendix, we present the shifted predictions obtained by relaxing the summation lower bound as indicated in Table \ref{Tab:sync}. A finite size scaling in a broad step length range could give us the exact expression.

\section{The experiments}\label{App: dialoguem}
\renewcommand{\theequation}{D-\arabic{equation}}
\setcounter{figure}{0}
\renewcommand{\thefigure}{D-\arabic{figure}}

%
%

As stated, the presented model is intented to simulate deliberatory spaces. This inspired the design of an instant messaging app based on Telegram, named "Dialoguem!" (meaning "Let's talk!" in Catalan)\footnote{https://github.com/Dialoguem/telegram-bot}. Dialoguem! can be view both as a research device to perform experiments, and as a device to assist deliberation in real settings. A series of ongoing pilot experiments aims to validate and potentially calibrate the model in its synchronized variant. In Dialoguem!, participants are presented with a question via a Telegram chatbot, and are encouraged to express their opinions anonymously within the model's defined opinion segment. Following this, they articulate a sentence to elucidate their stance. Subsequently, participants engage in reading and rating all other participants' sentences, then they assess their inclination to reach an agreement with the sender of each statement (i.e., if they fall within the model's interaction threshold). As the process unfolds, if participants modify their initial opinions, they adjust their position within the opinion segments. This iterative process continues until all participants confirm they do not wish to change their positions within the opinion segments any further. The evaluation of other participants' statements holds significant importance as self-assessment of opinions tends to be subjective and inconsistent. However, aggregating ratings from all participants regarding a given opinion yields more accurate predictions of shifts within the opinion segments between rounds. Nonetheless, it's noteworthy that the results of the experiment are still inconclusive from a statistical perspective.

\begin{figure}[h!]
\centering
\begin{subfigure}{0.49\linewidth}
    \centering
    \includegraphics[width=\linewidth]{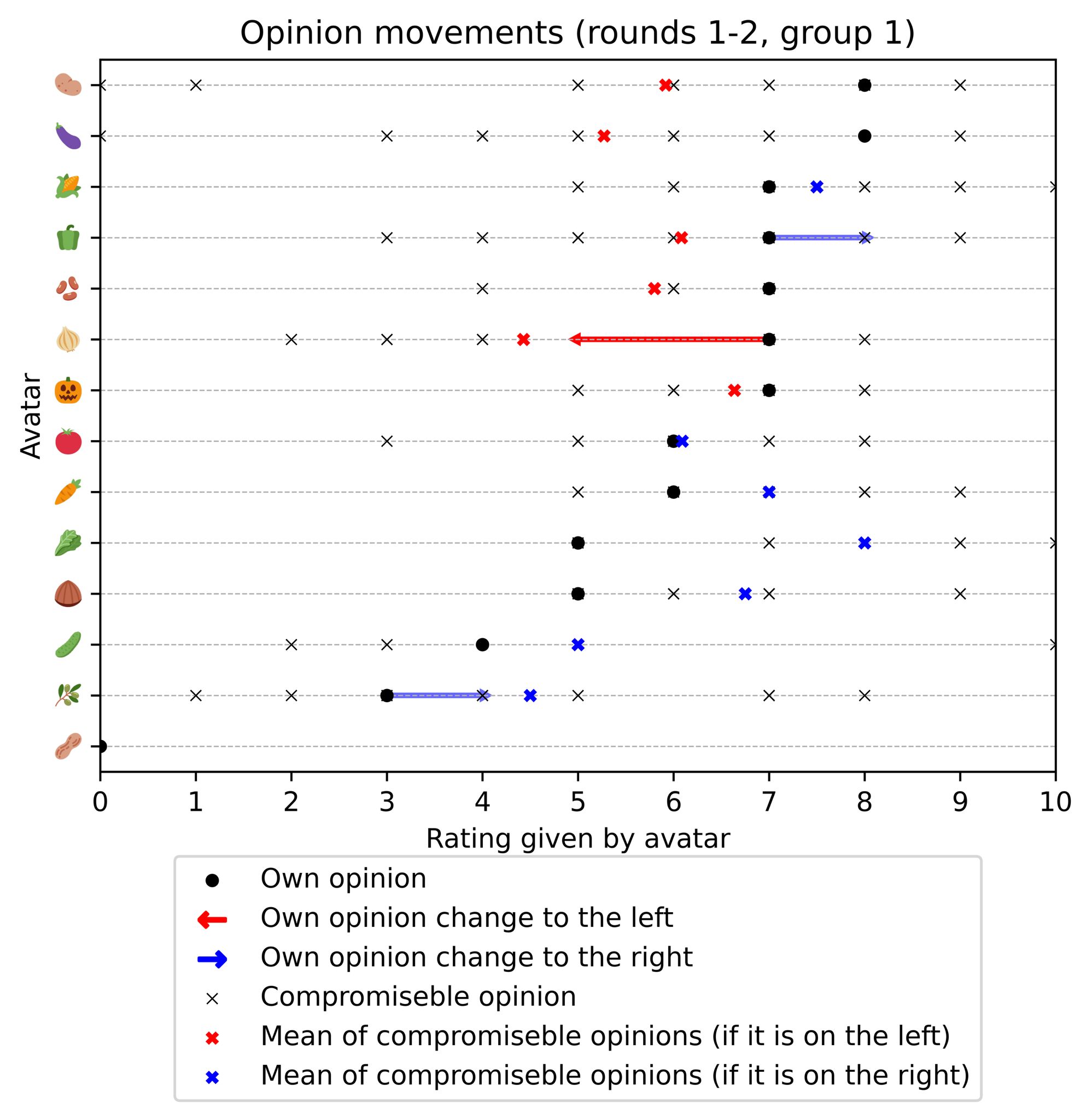}
    \put(-220,217){\textbf{(a)}}  
    \label{fig:subfig1}
\end{subfigure}
\hfill
\begin{subfigure}{0.48\linewidth}
    \centering
    \raisebox{0mm}{\includegraphics[width=\linewidth]{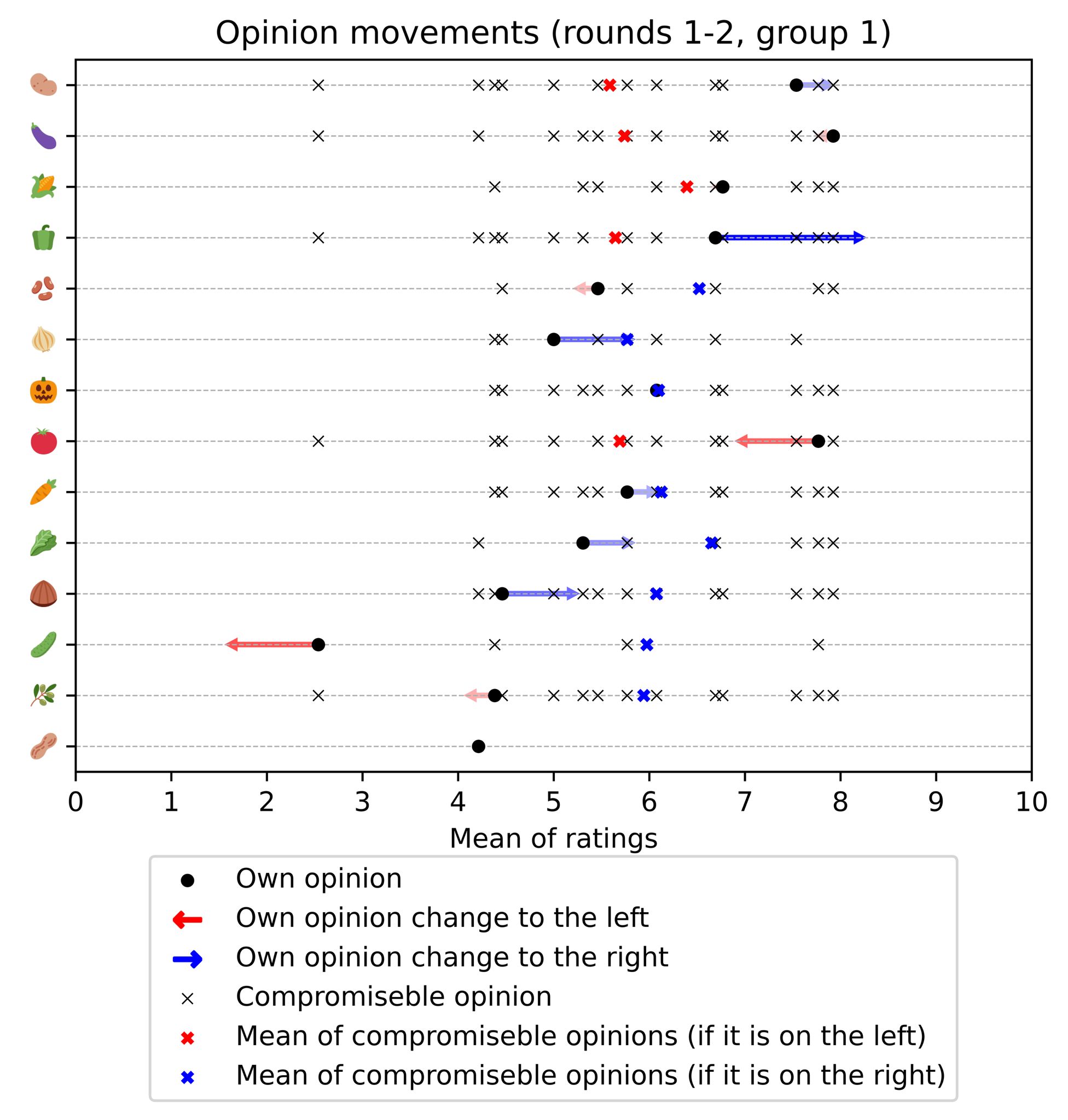}}
    \put(-220,217){\textbf{(b)}}  
    \label{fig:subfig2}
\end{subfigure}
\caption{Example of the participants' opinion positions and displacements between two rounds of the experiment: (a) given by each participant, represented by their avatar, and (b) as the mean rating given to each participant by all other participants. The right image shows a smaller dispersion of the opinions that each participant agreed to compromise on (grey crosses).}
\label{Fig:clust_sync_pairs_modified}
\end{figure}

\end{appendix}
\end{document}